\documentclass[10pt,conference,compsocconf,final]{IEEEtran}

\title{The Anatomy and Facets of Dynamic Policies}
\author{
  \IEEEauthorblockN{Niklas Broberg \qquad Bart van Delft \qquad David
    Sands}
  \IEEEauthorblockA{Chalmers University of Technology, Sweden}
}

\newcommand{\para}[1]{\par \noindent\textbf{#1}}

\usepackage{listings}
\usepackage{color, colortbl}
\definecolor{Gray}{gray}{0.9}

\usepackage{cite}

\usepackage{centernot}
\newcommand{\mayFlow}{\ensuremath{\longrightarrow}}
\newcommand{\mayNotFlow}{\ensuremath{\centernot\longrightarrow}}

\usepackage{silence}

\usepackage{amsmath,amssymb}
\usepackage{ifthen}
\usepackage{xargs}

\usepackage{flushend} 

\newcommand{\know}{{\it k}}
\newcommand{\ek}{{\it ek}}

\newcommand{\Sseq}{\overrightarrow{S}}

\newcommandx{\yaHelper}[2][1=\empty]{%
\ifthenelse{\equal{#1}{\empty}}%
  { \ensuremath{ \scriptstyle{ #2 } } } 
  { \raisebox{ #1 }[0pt][0pt]{ \ensuremath{ \scriptstyle{ #2 } } } }  
}   

\newcommandx{\yrightarrow}[4][1=\empty, 2=\empty, 4=\empty, usedefault=@]{%
  \ifthenelse{\equal{#2}{\empty}}
  { \xrightarrow{ \protect{ \yaHelper[ #4 ]{ #3 } } } } 
  { \xrightarrow[ \protect{ \yaHelper[ #2 ]{ #1 } } ]{ \protect{ \yaHelper[ #4 ]{ #3 } } } } 
}

\newcommand{\overarrow}[2]{\textnormal{\raisebox{-0.5ex}{$\yrightarrow{#1}[-1pt]\mathrel{\vphantom{\to}^{#2}}$}}}

\newcommand{\Rx}{\textsc{Rx}}

\usepackage{tikz}
\usetikzlibrary{arrows,automata}

\newtheorem{remark}{Remark}

\usetikzlibrary{positioning}

\usepackage{paralist}

\usepackage{rotating}
\usepackage{multirow}

\usepackage{framed}
\usepackage{hyperref}
\begin{document}

\maketitle

\begin{abstract}
Information flow policies are often dynamic; the security concerns of a program
will typically change during execution to reflect security-relevant events.
A key challenge is how to best specify, and give proper meaning to, such
dynamic policies. A large number of approaches exist that tackle
that challenge, each yielding some important, but unconnected, insight.
In this work we synthesise existing knowledge on dynamic policies, with
an aim to establish a common terminology, best practices, and frameworks
for reasoning about them. We introduce the concept of \emph{facets} to
illuminate subtleties in the semantics of policies, and closely examine
the \emph{anatomy} of policies and the \emph{expressiveness} of policy
specification mechanisms. We further explore
the relation between dynamic policies and the concept of declassification.
\end{abstract}

\section{Introduction}
Many of the security concerns that arise in software can be understood 
in terms of information flows.
For example, confidentiality requires that sensitive data does not flow to places
where it can be observed by unauthorised subjects, whereas integrity requires that
untrusted data does not flow to places which are trusted by other components in the system. 
Although information flow policies can express these concerns, in practice the
desired information flows that a system permits is not a static notion, but one
which may change during the running of the system to reflect security-relevant
events. For example, information under a restrictive information flow policy may
be less restricted if the information has been purchased. 
Conversely, the privileges of a principal may be revoked, thus reducing the
information flows deemed acceptable. We refer to information flow policies that
may change during a single instance of a system as \emph{dynamic information flow policies},
or simply \emph{dynamic policies}. 

Over the last twenty years a large number of systems and mechanisms presented answers to questions as ``How can we express policies that let us specify which flows are acceptable?'' and ``How can we express semantic properties that precisely and succinctly capture what it means for a program not to contain unintended leaks?''.
With this plethora of answers, another set of questions naturally arise: which are the
\emph{best} answers, in a given context or for a specific task? By which means can we
compare such systems and mechanisms to one another?
Which policy specification mechanisms are suitably expressive to capture the
security requirements of a specific use case? What proposed semantic property
gives strong enough guarantees against a particular kind of attacker?


To these new \emph{meta}-questions, we have far fewer answers. With a few
exceptions, what we have for guidance are \emph{case studies} that compare
systems in the light of one specific context or task 
\cite{Stoughton+:PLAS14,Askarov:Sabelfeld:ESORICS2005,Preibusch:POLICY2011,Hicks+:ACSAC06}.
While certainly useful and valuable for the insights they do give, case studies are
by nature not suited for drawing general conclusions or establishing
principles.

This paper also provides no distinct answers to these questions; rather it aims to 
give guidance to those who seek to answer them
by providing a synthesis of knowledge on dynamic policies.
Our over-arching aim and contribution is to provide
a clearer picture of information flow control in the presence of dynamic policies,
to facilitate understanding, defining, analysing, comparing, and discussing
properties and mechanisms. One of our key contributions, running as a theme through
the paper, is therefore to establish and argue for a \emph{common vocabulary},
and to identify \emph{common best-practices} in how we work with dynamic 
information-flow policies. 

Our study explores three main areas: how policies
are given semantics; a classification of the types
of dynamic policies that exist; and finally a reflection on the relationship
to the concept of declassification. We summarise the contributions of each of these in turn.

\para{Semantic Specification of Dynamic Policies} (\S\ref{sec:semantics}) 
What is the semantic meaning of a dynamic policy, and what are the best practices 
for specifying them?
To give a semantics to an information flow policy one must define a security property:
what does it means for a program to \emph{satisfy the given policy}? In other words, when
are the information flows that occur when executing a program permitted
by the policy?  When a program satisfies a given policy we say that the program
is \emph{secure} (with respect to that policy). But giving a semantics to a
dynamic policy can be subtle, complicated, and often unintuitive. It turns out,
as we will show, that small and seemingly minor changes to the details of a
security property can lead to fundamental
differences in which programs are considered semantically secure, differences
that were previously poorly studied and understood. 
To help put focus on these differences, we introduce the general concept of \emph{facets}
of dynamic security properties.
A facet captures a type of information flow which is permitted by some definitions but not by others.
We identify a number of such facets, argue for how they can be approached, and review how the literature 
treats these facets. 

\para{Classification of Dynamic Policies} (\S\ref{sec:pollang})
Dynamic policies can be specified in a variety of ways, offering a variety of expressiveness. 
In the second part of our synthesis we take a closer look at policy specification mechanisms.
We explain the general \emph{anatomy} of a dynamic policy in terms of a three-level
\emph{hierarchy of control}. 
Level 0 control refers to the ability of a policy mechanism to specify \emph{which flow relations} 
 can arise in a dynamic policy. 
Level 1 control refers to the way a mechanism controls \emph{how} the policy changes during execution.
Level 2 control refers to the meta-policy, i.e. how a mechanism controls \emph{which policy changes}  may occur.
From this hierarchy we derive a framework for formal comparisons of the expressiveness of policy specification
mechanisms, in terms of what \emph{invariants} they can enforce.
For example, can a given mechanism represent an invariant such as
``flows from top-secret to secret are never allowed''? Can a mechanism enforce that the permitted 
information flows always increase (or always decrease) over time? 
On top of our hierarchy we apply the ``dimensions of declassification'' introduced by Sabelfeld and Sands
\cite{Sabelfeld:Sands:JCS} 
to shed new light on the actual meaning and influence of these ``dimensions'' on dynamic policies. 

\para{Declassification} (\S\ref{sec:declassification})
Finally, we reflect on the concept of ``declassification'' and its relation to dynamic policies.
We look at two different flavours of declassification -- \emph{relabelling} and
\emph{copying release} -- and discuss how these can be interpreted and explained through
dynamic policies. Further we identify relationships between support for declassification, 
and the choices we make for certain facets of security properties.

\section{Terminology}

\label{sec:terminology}

Much of the terminology used in this paper is overloaded 
and used inconsistently across the literature. Here we fix the basic terminology
used in this paper, with the additional explicit purpose to establish a common vocabulary.


An \emph{information flow policy} refers to a specification of the information flows which are permitted
during program execution. At any given point, the permitted flows are given by a \emph{flow relation}.
This, for example, might be a specification that input variable $x$ is allowed to flow to output channel $y$. 
When the permitted information flows do not change over time, i.e. a single flow relation is used exclusively
throughout computation, we refer to the policy as a \emph{static policy}. 
When the permitted flows (may) change during computation, we call it a \emph{dynamic policy}.
A dynamic policy is thus a specification of a set of flow relations, any one of which is active at a
given point in time, together with a specification of how the system transitions between them.\footnote{Observe
the difference between a dynamic policy and a \emph{dynamic label}. The latter refers to a security label
which is only known at run time. The label itself however does not change during execution, neither does
its relation to other labels (see e.g.~\cite{zheng2005dynamic}).}

\newcommand{\kBefore}{\ensuremath{\mathcal{K}_{\it before}}}
\newcommand{\kNow}{\ensuremath{\mathcal{K}_{\it now}}}
\newcommand{\rNow}{\ensuremath{\mathcal{R}_{\it now}}}
\newcommand{\obs}[1]{\ensuremath{{\it obs}_A(#1)}}

Information flow policies are often specified indirectly via a set of \emph{labels}. For example, the classic static two-level security setting is defined by a set of two labels $\{L, H\}$ (for low and high confidentiality, respectively). A \emph{flow relation} is then described by (i) assigning labels to appropriate sources and sinks in a configuration, and (ii) defining a relation between labels, indicating allowed flows.  For example in the classic two-level setting, secret inputs are labelled $H$, and public outputs are labelled $L$, and the flow policy is specified by saying that the may-flow relation is the smallest reflexive relation such that $L$ may flow to $H$, which we write as $L \rightarrow H$. 

A \emph{policy scheme} is a set of labels, their flow relations and transitions, viewed in isolation from the label
assignment in any particular program or system. In a policy scheme, a flow relation therefore only consists of the
relation between labels (point (ii) above). Finally, different approaches provide different mechanisms (languages)
with which to construct such policy schemes, and different ways to assign labels to relevant entities. We refer to
these as a \emph{policy specification mechanism}.




\section{Semantics of Security Properties}
\label{sec:semantics}



Non-interference is the name usually given to the semantic definition 
of when a program (or system) satisfies a
\emph{static} policy. In the absence of subtleties arising
from non-determinism and interaction, non-interference could be argued
to be a simple and intuitive
property, easily stated in the following or any one of several equivalent forms:
If information from input $i$ is not allowed to flow to output $o$,
then no variety in $i$ may result in a different output $o$~\cite{Cohen:InformationA}.

Even in the settings where there exist various forms of
non-interference, they are often related according to various trade-offs
between security and permissiveness \cite{Zakinthinos:SP97,Focardi:Gorrieri:Classification}.
However, once we abandon static
policies the situation changes dramatically.  Not only does each new mechanism for
specifying dynamic policies introduce a new semantic
definition, the definitions are 
typically formulated in completely new
styles with different attacker models. How these models relate to 
one another is not easy to see even with extensive scrutiny.
Nor is it intuitively clear at first glance just what notion of security a particular
property actually guarantees.


In the case of dynamic policies we believe that 
 different systems have more fundamentally different semantic requirements, and
different notions of security and attacks. Our grand quest then changes,
from one of finding \emph{the} security property, to that of identifying
how best to state a security property for a
specific set of requirements.

In this section we discuss various aspects of the art of defining
information flow security properties. We do not propose any new
security properties, instead we identify and illuminate relevant principles
and \emph{facets} that affect such properties, and suggest practical consequences
and interpretations of the various choices that can be made. Furthermore,
we give a survey of previously proposed properties from other work and show how
they fall within the spectrum of our facets.

Our purpose here is two-fold: firstly, we hope to provide tools for comparing
and reasoning about security properties, to more easily understand how
they relate or differ. Secondly, we want to provide the would-be property
author with a toolbox of best practices and relevant concerns, to assist
in stating a property suitable for that author's specific context.

This section will proceed as follows: First, we argue strongly for why
so-called knowledge-based, or \emph{epistemic}, formulations of properties
are preferable to the traditional two-run-style formulations. Attempts
at such arguments have been made previously
\cite{Askarov:Sabelfeld:Gradual, askarov2012, Broberg:Sands:PLAS09},
but in passing and never with a complete and coherent picture.
Second, we identify a number of \emph{facets} of information
flow security properties. We give illustrating examples, and discuss
how previously proposed semantic properties differ, sometimes subtly,
in the choices made. As we will show, typically no universally ``best''
choice exists for these facets, which further emphasises the futility
of trying to find a single property that works for all situations.
Finally we categorise existing work according to the facets we identify.

\subsection{Epistemic information flow security}
\label{sec:semantics:epistemic}

Askarov and Sabelfeld introduced the \emph{Gradual release} property
\cite{Askarov:Sabelfeld:Gradual} for information flow control using an explicit
model of \emph{attacker knowledge evolution} (also called an
\emph{epistemic} property) when observing a single program run. 
Askarov and Sabelfeld
drew inspiration from work on \emph{deducibility} \cite{Askarov:Sabelfeld:Gradual},
which in turn borrowed the technique from work on possibilistic security
\cite{Zakinthinos:SP97,Halpern:2008}.
Some later systems have used epistemic formulations
\cite{Broberg:Sands:PLAS09,askarov2012,Balliu2013,Banerjee+:Expressive}.

The main purpose of this section is to argue that epistemic
formulations are attractive in that they capture the desired property in a natural and direct fashion. 
We wish to be clear that the definitions in this section are not essentially novel, but generalised versions of those presented in the cited literature.


The core feature of an epistemic formulation of security is that we consider the system from the perspective of an attacker's knowledge of that system, and how observations on the system change this knowledge.
%
For some observation, we refer to the information that is allowed to be revealed at the moment that observation is produced as the \emph{release policy} (terminology cf. Balliu~\cite{Balliu2013}), denoted \rNow{}, and to the attacker's knowledge just before and after this observation as \kBefore{} and \kNow{}.
With these simple terms we can present the general form of an epistemic security property:
\[
  \rNow \text{ allows } {\it increase}(\kBefore, \kNow)
\]

We now continue by specifying a concrete computational system and what is meant with knowledge, although different models could have been used.
We say that the execution of a program yields a \emph{trace} of program \emph{events}.
We deliberately remain abstract in the nature of the events themselves. 
For a program $P$ and an initial memory store $S$, we denote with $\langle P, S \rangle \overarrow{t}{}$ that $t$ is a prefix of the event trace produced by $P$.
Let \obs{t} be a function determining what observation the attacker $A$ makes on the event trace $t$.
Again we remain abstract to what these observations are, but options include the series of values output on $A$'s channel and updates to memory locations observable by $A$.
It also possible to have more complicated observation functions, an example of which can be found in the work by Askarov and Chong~\cite{askarov2012}.

We model the inputs to the system as the initial memory.
The attacker's knowledge can thus be expressed as how much the attacker knows about these initial values in the memory, based on the observations made.
In existing work such as Gradual Release, this is presented as the set of initial memory stores for which the program can produce the observation made by $A$.
That is, if $\langle P, S \rangle \overarrow{t}{}$ with $\obs{t} = o$, the attacker $A$'s knowledge on $S$ is:
\[
  k_A(P,o) = \{ S' | \langle P, S' \rangle \overarrow{t'}{} \text{ and } \obs{t'} = o \}
\]
To arrive at a more natural definition of knowledge \emph{increase} we follow the example of van Delft et al. \cite{delft2015} and work instead with the complementary notion of \emph{exclusion knowledge} -- the set of initial memories that could \emph{not} have led to the observed trace, i.e. the set complement of $\know_A$:
\[
  \ek_A(P,o) = \overline{\know_A(P,o)}
\]
Knowledge gain can then conveniently be expressed as the set of stores that the attacker \emph{additionally} excludes after making a next observation.
Let $\langle P, S \rangle \overarrow{t \cdot e}{}$  denote that event $e$ was preceded by the trace of events $t$.
$A$'s increase in knowledge is described as the difference between 
$A$'s knowledge before and after event $e$:
\[
  \ek_A(P,\obs{t \cdot e}) \setminus \ek_A(P,\obs{t})
\]
where $\setminus$ is the set difference operator.
Finally, the release policy that specifies the \emph{permitted} increase in knowledge can be expressed 
as an \emph{upper bound} on the knowledge gained:
\[
  \ek_A(P,\obs{t \cdot e}) \setminus \ek_A(P,\obs{t}) \subseteq \rNow
\]
where $\rNow$ is also a set of memory stores.
For example, in the setting of two-level non-interference we want to forbid the attacker from learning anything about the secrets in the initial store.
Hence, non-interference can be encoded by making $\rNow$ the set of stores that have different values from the actual initial store $S$ on the locations not containing secrets.

Depending on the policy language used, $\rNow$ can be parameterised over a wide range of aspects -- in particular the attacker $A$ and the flow relation dictated by the dynamic policy when $e$ was produced.
Other possible aspects include the event $e$ itself, the trace $t$, the current
memory store at the time of the observation, the current program at the time of the
observation, or some entirely external entity, for example the system clock.

In words, this abstract property can be expressed as ``an attacker observing an
event produced by a running program cannot learn anything about the initial memory
that is not allowed by the policy at the point of the event''.

We posit that this very simple abstract formulation captures the vast majority of properties
we want to express through different
instantiations of observation models $\obs{\cdot}$, release policies $\rNow$ and observable traces of
events $t \cdot e$. But while this abstract form is quite simple,
instantiating it properly can be quite tricky.

\para{Why epistemic?}
Consider a two-run formulation of non-interference for a deterministic program. 
If spelled out in words, it
would read something like 
(disregarding termination): 
``if the program is run twice with the same public inputs but possibly different secrets, 
the public outputs must be the same in both runs''. 
We treat the concepts of inputs, outputs and the notion of public informally here,
what is important is that
the above quote conveys the gist
of such a two-run formulation. Now consider how one would convince a non-expert
that this is indeed a suitable characterisation of a program that does not
leak any secret information. That argument would very likely be something like:
``if this holds, then an attacker observing the outputs of
running the program could not deduce anything about what the secret inputs are.''
But that is exactly what the epistemic property states! The notion of security
intrinsically has nothing to do with observing two separate runs -- but rather
what can be deduced from observing a single run.

The above argument strongly suggests that the epistemic formulation is the most
natural way we can state the desired property. A two-run formulation could certainly
be very useful as part of the strategy to \emph{prove} e.g. 
the correctness of an enforcement mechanism; often
a (mechanical) two-run formulation can lend itself well to the structure of a proof
over execution traces. But that property is then only a stepping stone, and should,
for completeness, be shown to imply the natural epistemic property.

Another, purely technical reason to prefer epistemic formulations over \emph{bisimulation-based}
two-run properties is pointed out by \cite{Broberg:Sands:PLAS09}: the latter are often overly conservative.
We return to this point in section \ref{sec:semantics:survey}.


\subsection{Facets of Semantic Security Properties}
\label{sec:semantics:dimensions}

When defining an information flow security property, choices are made that ultimately affect what programs are considered secure according to that property.
These choices determine what we call the \emph{facets of security properties}. 

\begin{framed}
\centerline{\bf Facet}
\noindent
An aspect of a security property that determines whether a particular class of information flows is accepted as secure.
\end{framed}

{\bf Note:}
The aim of this section is not to formally define these particular classes of information flows.
Rather, we point out several examples that underline the existence of facets as a design space for security properties.

Various effects could be regarded as facets of a security property.
A well-known facet is \emph{termination sensitivity}: a property can either allow programs to leak sensitive information through their termination behaviour, or not.
This is a facet of which current designers of security properties (and
enforcement mechanisms) are well aware.
A conscious choice is made if this facet is addressed, often
motivated from the pragmatic perspective that an enforcement for a
termination insensitive property is easier to achieve and less
restrictive. 
%
More facets of security properties exist, but their existence is largely unknown and hence how a property treats these facets is not by conscious choice.

In this paper we are interested in those facets that arise in various semantics properties of dynamic policies. 
Each facet is not universal among the works that we have surveyed --
some definitions permit a facet while at least one other does not. But in most cases
the \emph{choice} for each facet is far from explicit. 
We attempt to present a justification for both the permissive as well as the restrictive treatment of each facet, although arguably these pleas are not equally compelling in all cases.

Our main contribution of this section is \emph{not} to identify these facets -- some facets were identified before, and we make no claim that we have identified all facets -- but to argue for recognising the \emph{existence} of facets and provide the designer of a security condition with the background to make informed choices for them.

The facets that are discussed in this paper are \emph{time-transitive flows}, \emph{replaying flows}, \emph{direct release} and \emph{whitelisting flows}.
We phrase each of these facets in terms of what it means to consider these flows as secure, and attempt to present the same flow in two contexts, each motivating whether the flow should be considered secure or not.

{\bf Note:} It is important to point out that the (in)security of these examples is argued only using the code fragments and potential application context, not in the view of any particular security condition.
This underlines the principle we aim to convey: when designing a security condition, first decide how the condition should treat these facets and then construct the right condition to match those choices, not vice-versa.

\begin{remark}[Notation]
  In our examples we maintain the convention that program locations are labelled with a fixed 
security level.
	The lower-case first letter of the program location matches the upper-case first letter of the security level (e.g. \lstinline!a! has level \lstinline!A! and \lstinline!hos! has level \lstinline!Hospital!).
	We assume that initially no information flows are allowed between any two levels.
	The syntax \lstinline!B! $\mayFlow$ \lstinline!A! changes this ordering and allows information to flow from level \lstinline!B! to \lstinline!A!, whereas \lstinline!B! $\mayNotFlow$ \lstinline!A! revokes this permission.
	Note that these are only conventions used to present the flows; the flows themselves do not rely on them. This notation is taken from \cite{askarov2012}.
\end{remark}

\para{Time-transitive flows}
\emph{A flow is time-transitive if it moves information from security level $A$ to level $C$ via a third level $B$, while there is no moment in time where the flow from $A$ to $C$ itself is allowed by the flow relation.}

\begin{center}
\begin{tabular}{|p{.45\hsize}|p{.45\hsize}|}
\hline
\rowcolor{Gray}
  {\bf\small Secure} & {\bf\small Insecure} \\
\begin{lstlisting}
  User *\mayFlow* XSSFree
x := escapeHTML(uIn);
  User *\mayNotFlow* XSSFree
  XSSFree *\mayFlow* DB
db := escapeSQL(x);
\end{lstlisting} &
\begin{lstlisting}
  Patient *\mayFlow* Hospital
hos := patData;
  Patient *\mayNotFlow* Hospital
  Hospital *\mayFlow* DrPhil
drPhil := hos;
\end{lstlisting} \\ \hline
\end{tabular}
\end{center}

In a context where the flow of information reflects some declassifying or sanitising intention, one can argue that time-transitive flows are secure.
In this example, user input \lstinline!uIn! is first passed through the sanitiser \lstinline!escapeHTML! to prevent XSS attacks, and later through the sanitiser \lstinline!escapeSQL! before storing the information in the database.
Here, it is of no relevance that the user input was never allowed to flow directly to the database.

The time-transitive flows facet was previously identified under the name \emph{transitive flow} by Swamy et al.\ in the development of \Rx{}~\cite{Swamy+:Managing}.
We purposely refer to them as time-transitive flows to avoid confusion with intransitive non-interference~\cite{Rushby:92}, which is discussed further in Section~\ref{sec:pollang:reclassify}.
Swamy et al. argue that time-transitive flows should be considered insecure, using the following example.
Patients allow their data to only flow to the doctors of the hospital while they are under treatment.
When a patient leaves the hospital, this information should no longer be available and in particular not to doctor Phil, who joined the hospital staff after the patient has left.
Here, it is sensible to disallow the time-transitive flow.

One way to differentiate between these in the security property is to either limit the attacker's increase in knowledge by what can be learnt from the observable part of the \emph{current} memory (allowing time-transitive flows)\footnote{This can be translated into knowledge on the initial memories that could have resulted in the current observable memory.} or from the \emph{initial} memory (disallowing time-transitive flows).

\para{Replaying flows}
\emph{When the release of information is considered \emph{permanent}, this information flow can be repeated without breaking the information security of the system.}

As an example, consider the scenario where the National Security Agency (NSA) releases a file to the US military.
Once released, the military can access this file at any time, regardless of whether this information is currently in their possession.

\begin{center}
\begin{tabular}{|p{.45\hsize}|p{.45\hsize}|}
\hline
\rowcolor{Gray}
  {\bf\small Secure} & {\bf\small Insecure} \\ 
\begin{lstlisting}
  NSA *\mayFlow* Military
mil := nsaFile
mil := 0;
  NSA *\mayNotFlow* Military
mil := nsaFile;
\end{lstlisting} &
\begin{lstlisting}
  Creditcard *\mayFlow* Log
log.write(cc);
log.clear();
  Creditcard *\mayNotFlow* Log
log.write(cc);
\end{lstlisting} \\ \hline
\end{tabular}
\end{center}

Considering information as permanently released is not the natural choice in every situation, as the insecure example demonstrates.
Since the log file has been cleared, the credit card information is no longer available.
Hence the effect of the earlier release has disappeared, and to store the same information in the log file again requires the flow relation to agree with this flow again.

To make this difference even more explicit, we include a second example for both the secure and insecure context which combines replay with time-transitive flows.
If the NSA information has been permanently released to the military, when Bob later joins the military he should have access to this information as well, again regardless of whether it is currently in the military's possession.
In the insecure context, the vendor is allowed to see the information in the log file, and since it does not contain the credit-card number, the vendor should not be allowed to observe it.

\begin{center}
\begin{tabular}{|p{.45\hsize}|p{.45\hsize}|}
\hline
\rowcolor{Gray}
  {\bf\small Secure} & {\bf\small Insecure} \\ 
\begin{lstlisting}
  NSA *\mayFlow* Military
mil := nsaFile;
mil := 0;
  NSA *\mayNotFlow* Military
  Military *\mayFlow* Bob
bob := nsaFile;
\end{lstlisting} &
\begin{lstlisting}
  Creditcard *\mayFlow* Log
log.write(cc);
log.clear();
  Creditcard *\mayNotFlow* Log
  Log *\mayFlow* Vendor
vendor.receive(cc);
\end{lstlisting} \\ \hline
\end{tabular}
\end{center}

The restricting interpretation appears more natural when taking a language-based perspective on information release.
The view of permanently releasing information matches more closely the original use of the term ``declassification'' in a military context, whereas the language-based approach is more related to how the same term ``declassification'' is often used in current information flow research.
We return to this disambiguation of the term declassification in Section~\ref{sec:declassification}.


The language-based view suggests that we can make a second distinction in this facet, which we call \emph{weak replaying} of flows.
Weak replay captures the idea that information is only considered released as long as it is still available at the level to which it was released. 
A motivating example for weak replays is again the setting where the same credit card information is added to the log file, but before the log has been cleared.

\begin{center}
\begin{tabular}{|p{.45\hsize}|p{.45\hsize}|}
\hline
\rowcolor{Gray}
  {\bf\small Secure} & {\bf\small Insecure} \\ 
\begin{lstlisting}
  Creditcard *\mayFlow* Log
log.write(cc);
  Creditcard *\mayNotFlow* Log
log.write(cc);
\end{lstlisting} &
\begin{lstlisting}
  Ezine *\mayFlow* Customer
customer := ezine;
  Ezine *\mayNotFlow* Customer
customer := ezine;
\end{lstlisting} \\ \hline
\end{tabular}
\end{center}

To argue for the insecurity of weak replaying flows, consider a scenario where a customer pays for a time-limited subscription
on an online magazine (``e-zine''). When the subscription runs out, the customer should no longer be able to download
magazine, even if they have an old copy of the same edition. 


This facet was previously identified by Askarov and Chong, and we use their approach as a technique for addressing this facet in its various degrees~\cite{askarov2012}.
To allow for (strong) replaying of flows, we can set the attacker's observation power to remember all observations made.
That is, $\obs{t}$ could be said to be the sequence of events observable by $A$.
Hence, after observing an event $e$ which contains the same information flow in earlier observations $\obs{t}$, we have that $\ek_A(P,\obs{t \cdot e}) = \ek_A(P,\obs{t})$ and the release is considered secure regardless of the current flow relation.
To disallow any replaying of flows, we can consider attackers who do not have a perfect recall of all observations made, and to whom the replay may therefore come as a revelation.

Again, our aim is only to argue for the existence of facets, not to state that one treatment of a facet should be preferred over another.
For replaying flows we see examples of both choices in the reviewed literature in Section~\ref{sec:semantics:survey}.

Although allowing for (only) weak replaying flows seems to arguably better match the language-based 
view, we are not aware of any literature that addresses the facet in exactly this way.
One possible encoding  would consider attackers without perfect recall, but allow the attacker to observe the (currently) non-secret part of the current memory.

\para{Direct Release}
\emph{A security condition supports direct release if information is considered released as soon as the current flow relation permits it to flow.
This means that a revocation of that permission does not affect this information.}\footnote{Note that despite the similarity in our examples,
direct release is not merely an even stronger version of replaying flows. Direct release is concerned with what an attacker is assumed to
have learned everything that is permitted at the point when the flow relation becomes more liberal -- not whether the attacker may learn it again once the flow relation no longer explicitly permits it.}

\begin{center}
\begin{tabular}{|p{.45\hsize}|p{.45\hsize}|}
\hline
\rowcolor{Gray}
  {\bf\small Secure} & {\bf\small Insecure} \\ 
\begin{lstlisting}
  Data *\mayFlow* App
send(app, "Hello");
  Data *\mayNotFlow* App
send(app, data);
\end{lstlisting} &
\begin{lstlisting}
  Salary *\mayFlow* Screen
screen.show("Hello");
  Salary *\mayNotFlow* Screen
screen.show(salary);
\end{lstlisting} \\ \hline
\end{tabular}
\end{center}

Considering such flows secure can be justified if we model attackers as constantly observing, directly in memory, 
all the information which they have permission to know.
As an example, the attacker could be an application running in parallel with the code displayed above.
Hence it does not matter that no data was actively sent to the application, we consider the data as released directly when this is allowed.
Note that direct release does not imply that revocation (changing the policy to be more restrictive) is irrelevant: new information that 
arrives at level \lstinline!Data! (either via input channels or from a different security level) is considered to be not yet released to \lstinline!App!.

On the other hand, we can argue that the same kind of flow is insecure when the attacker can only observe information that is actively provided.
In the code above no information about the salary has been released to the screen, and hence it makes sense to assume that an observer does not 
know this information yet.

If we chose to allow direct release, we could reflect this in the security property by modelling an attacker's observation as the part of the 
current memory that the attacker is allowed to observe according to the current flow relation.
This opposed to only considering ``active'' flows, such as observing changes in the memory, which we could use if we want to consider direct release insecure.

\para{Whitelisting flows}
\emph{A security property is whitelisting if a flow is allowed whenever there is some part of the policy that allows for it.}
This opposed to blacklisting, where a flow is disallowed whenever there is some part of the policy that does not allow it.
The facet becomes apparent when a flow is permitted by one part of the policy, but denied by the other.

As an example of such a situation, and an argument for whitelisting, consider the release of an encryption key.
It is reasonable to accept that with the release of this key an observer also learns the secret information that was earlier released encrypted under that key, even though part of the policy does not allow the secret to be released.

\begin{center}
\begin{tabular}{|p{.45\hsize}|p{.45\hsize}|}
\hline
\rowcolor{Gray}
  {\bf\small Secure} & {\bf\small Insecure} \\ 
\begin{lstlisting}
  Secret *\mayFlow* Pub
  Key *\mayFlow* Pub
output(k XOR secret);
  Secret *\mayNotFlow* Pub
output(k);
\end{lstlisting} &
\begin{lstlisting}
  Bob *\mayFlow* Report
  Carla *\mayFlow* Report
r.avg := (b.s+c.s)/2;
  Carla *\mayNotFlow* Report
r.bob := b.s
\end{lstlisting} \\ \hline
\end{tabular}
\end{center}

The insecure example shows the creation of a report \lstinline!r!, storing the average of the salaries \lstinline!s! of Bob and Carla.
Then, when Carla explicitly no longer allows information about her salary to flow to the report, we add Bob's salary to the report from which an observer can derive Carla's salary.
This we could argue violates Carla's concern and should be regarded as an illegal flow.

Whitelisting appears to be the norm for language-based security conditions as is confirmed by the literature that we discuss in this paper, which all treat the policy as a whitelist of permitted flows.
Treating the policy as a blacklist rather than a whitelist is more common in the interpretation of noninterference in event-based systems~\cite{Goguen:Meseguer:Noninterference,Zhang2012}.

\subsection{Classification of facets in literature}
\label{sec:semantics:survey}

\setlength\tabcolsep{5pt}

\begin{table}
  \centering
  \begin{tabular}{@{}l|l|*{4}{c}|@{}}
    \cline{3-6}
    \multicolumn{2}{ c|  }{} & T & R & D & W  \\ 
    \cline{2-6}
		\multirow{2}{*}{\rotatebox{90}{NI}}
    & Swamy et al. (\textsc{Rx})~\protect\cite{Swamy+:Managing}
      & + & - & + & + \\
    & Hicks et al.~\protect\cite{Hicks+:Dynamic}
      & + & - & - & + \\ 
    \cline{2-6}
		\multirow{2}{*}{\rotatebox{90}{BI}}
    & Boudol and Matos (Non-disclosure)~\protect\cite{Boudol:Matos:On}
      & + & - & - & + \\
    & Broberg and Sands (Flow Locks)~\protect\cite{Broberg:Sands:ESOP06}
      & + & - & - & + \\ 
    \cline{2-6}
		\multirow{5}{*}{\rotatebox{90}{Epistemic}}
    & Askarov and Sabelfeld (Gradual Release)~\protect\cite{Askarov:Sabelfeld:Gradual}
      & N/A & + & - & + \\			
    & Banerjee et al. (Flowspecs)~\protect\cite{Banerjee+:Expressive}
      & N/A & + & - & + \\
    & Balliu~\protect\cite{Balliu2013}
      & +/- & + & +/- & + \\
    & Askarov and Chong~\protect\cite{askarov2012}
      & - & +/- & - & + \\
    & Broberg and Sands (Paralocks)~\protect\cite{Broberg:Sands:PLAS09,Broberg:Sands:Paralocks}
      & + & + & - & + \\
	  \cline{2-6} 
		\multicolumn{6}{c}{}\\[-5pt]
		\multicolumn{6}{c}{T: time-transitive, R: replay, D: direct release, W: whitelisting}
  \end{tabular}
  \caption{Classifying existing security conditions along the facets.
    +~indicates that flows of this facet are allowed, -~that they are not.
    +/-~signifies that the facet is not fixed by the condition.
    N/A~denotes that the facet does not apply to this security condition.
		Grouped by nature of condition: non-interference (NI), bisimulation (BI)
		or epistemic.}
	\label{table:classification}
\end{table}\vspace*{0.7ex}

In this section we present a collection of existing security definitions for dynamic policies and classify them along our facets.
Although we think that our discussion is rather complete, our goal is not to give a full survey of the field.
Rather, the purposes of this section are
\begin{inparaenum}[\itshape a\upshape)]
\item to illustrate that the listed facets indeed occur in literature; 
\item to identify what components in a security condition determine which classification, as an aid for the developer of new security conditions;
\item to demonstrate that the facets can be used as a terminology for discussing security conditions for dynamic policies (similar to the intention of Sabelfeld and Sands introducing the dimensions for declassification~\cite{Sabelfeld:Sands:CSFW05}); and
\item to convince the reader that the intended facets should be considered before defining a security condition, to ensure that it matches these intentions.
\end{inparaenum}

With this setting in mind, we purposefully left out some literature that could be considered related to dynamic policies, but does not fit the goal of this section.
Examples include the \emph{non-interference until} conditions property from Chong and Myers~\cite{Chong:Myers:CCS04} (since it is unspecific in its treatment of information once released) and $\lambda$AIR by Swamy et al.~\cite{Swamy:2008} (which presents a type system for enforcing user-specified security properties, but does not present such properties itself).

An overview of the security conditions in the surveyed literature along our facets is listed in Table~\ref{table:classification}.
We remind the reader that the facets do \emph{not} serve as an objective measure for the quality of a security condition: arguments can be made both in favour and against each facet as illustrated in Section~\ref{sec:semantics:dimensions}.
Since none of the discussed literature interprets the policy as a blacklisting of information flows rather than a whitelisting, we omit this facet from this discussion.
Similarly, direct release is not permitted by most properties by virtue of their observation model and we only discuss this facet for the properties that do permit it.

\begin{remark}[Disclaimer]
  We present all conditions in the same computational model, despite some of the surveyed conditions being defined in a different context.
	As a consequence, some of the classifications in Table~\ref{table:classification} do not directly apply on these original security properties.
	For example, time-transitive flows do not apply on a property for a model where observations are made on output channels instead of state changes.
	We allow ourselves to make the following transformations: 
	\begin{itemize}
	  \item \emph{Modified sources}: We define sensitive information as the information in the initial state, instead of values input on channels.
	  \item \emph{Modified sinks}: We define observations as state modifications, instead of values output on channels.
	\end{itemize}
	We acknowledge that restricting ourselves to this single model may not be sufficiently general, and additional facets may be revealed under a different model (e.g. considering input channels instead of initial states, cf. Clark and Hunt~\cite{Clark:Hunt:FAST08}).
\end{remark}

We here only discuss the conditions to the extent needed to understand their classification.
We group the security conditions according to whether they are of a non-interference, bisimulation, or epistemic nature.

\subsubsection{Non-interference (NI) based conditions}

These conditions are built directly on top of non-interference properties phrased in two-run style.
Dynamic policies are enforced by varying how non-interference is included as a building block in the security condition, not by changing the non-interference property itself.

\paragraph{Swamy et al. (\textsc{Rx})~\cite{Swamy+:Managing}}
With the clearly stated goal of disallowing time-transitive flows (referred to as \emph{transitive flows}) \Rx{} introduces the notion of a \emph{transaction}.
Transactions are defined together with a fixed flow relation on parts of the information in the system.
If this specified flow relation is modified during the transaction, the operational semantics rolls back to the pre-state of the transaction, undoing the changes to memory but preserving the changes in policy.
This operational guarantee allows the static check to assume the fixed flow relation to hold throughout the transaction.

Although the goal is clearly stated to prevent time-transitive flows, they are only prevented by \Rx{} if the programmer explicitly states this intent by placing the program in a transaction with the right flow relation.
What is more, the security condition itself also does not rule out these flows which suggests that the condition does not match the intended treatment of this facet.

Rephrased in our computational model, the essence of the security condition \emph{non-interference between declassifying policy change} can be defined follows.\footnote{The actual definition and configuration more involved as they explicitly list specific elements such as the fixed flow relation. However, these are differences are not relevant for the classification of the security property along the facets.}
A memory store $S$ consists of both a regular variable mapping and a policy component describing the current ordering between the security levels.
From a configuration $\langle P, S \rangle$, an attacker $A$ observes the part of $S$ that according to the policy component in $S$ may flow to $A$.
We write $S|_A$ to denote the projection that $A$ observes.
Each step in the computation produces as event the current store; hence we denote with $t|_A$ the observation of $A$ on trace $t$, i.e. each store projected to $A$ as per the policy component in that store.
Given two stores $S_1$, $S_2$ such that $S_1|_A = S_2|_A$, producing two traces $\langle C, S_1 \rangle \overarrow{t_1}{}$ and $\langle C, S_2 \rangle \overarrow{t_2}{}$ such that each trace either terminates or the next step is a declassifying policy change (i.e.\ additional memory locations in $S$ become visible to $A$).
Then $P$ is non-interfering if $t_1|_A \mathop{\dot{=}} t_2|_A$ (where $\dot{=}$ is equivalence up to stuttering).
The end-to-end security property is then obtained by applying non-interference piece-wise to each non-declassifying sub-trace.

Until the non-interference condition is restarted the property considers only those stores that appear equivalent to $A$ according to the initial flow relation.
After the condition is restarted, the revoked flows require to consider stores that vary in information that may have flown previously.
For example, in the following program, the non-interference condition assumes that the value in both runs for variable $b$ is equal for the second assignment, but does not do so when restarted before the third assignment:
\begin{lstlisting}
  B *\mayFlow* A
a := b;
  B *\mayNotFlow* A
a := b;
  C *\mayFlow* A
a := b;
\end{lstlisting}
Hence the security property only allows replay of information until declassifying policy change, which we argue is effectively not supporting replay of flows.

Interestingly, although disallowing time-transitive flows was the motivation behind the transaction system, the security property does \emph{not} rule them out.
Consider the time-transitive flow of information from security level $C$ to $B$, followed by $B$ to $A$, as seen by an observer on level $A$:
\begin{lstlisting}
  C *\mayFlow* B
b := c;
  C *\mayNotFlow* B
  B *\mayFlow* A
a := b;
\end{lstlisting}
Since allowing the flow from $B$ to $A$ is regarded as a declassifying change for $A$, the non-interference condition restarts.
Consequently the security condition only considers memory stores that agree on all values of level $B$ for this second part of the flow, ruling the program as secure.
\Rx{} thus serves as an example where a facet of the security property clearly deviates from the intention.

Observations are modelled as projections of what the attacker may observe from the current memory under the current flow relation.
Revocation thus does not affect the classification of released information, making the security condition allow for direct release.

\paragraph{Hicks et al.~\cite{Hicks+:Dynamic}}

The Decentralized Label Model (DLM) is a well-established language for specifying information flow security labels based on principals~\cite{ML00}.
The ordering between security labels is partly influenced by an \emph{acts-for} hierarchy between these principals, which may vary over time.
Jif, an extension to Java adding support for the DLM, assumes that this acts-for hierarchy changes only very occasionally.
This justifies that programs can make queries on the hierarchy which are then assumed to hold, ruling the following program secure~\cite{Myers:POPL99}:

\lstset{
  basicstyle=\ttfamily\footnotesize,
  aboveskip=4pt,belowskip=2pt
}
\begin{lstlisting}
if (A *\mayFlow* B) { long_computation(); b := a; }
\end{lstlisting}

The work by Hicks et al. introduces a calculus that removes this assumption.
Coercion checks (permission tags) are statically introduced to summarise what constraints on the acts-for hierarchy need to hold in which parts of the program to be secure.
When at run-time an (asynchronous) request for changing the acts-for hierarchy presents itself, this is delayed until this change does not break the currently necessary permission tags.

A program is said to be secure if it is non-interfering for an unknown, but fixed, acts-for hierarchy.
Effectively, this results in the property \emph{non-interference between policy updates}.
%
To classify this condition along our facets, the flows need to be rephrased to use policy queries in a way that matches the appropriate policy changes.
For example, a replay of flow can be exemplified by:
\begin{minipage}{\columnwidth}
\begin{lstlisting}
   A *\mayFlow* B
if (A *\mayFlow* B) { b := a }
   A *\mayNotFlow* B
b := a
\end{lstlisting}
\end{minipage}
This example shows that the security condition does not allow the replay, since this program is interfering for a fixed policy in which \lstinline!A! may not flow to \lstinline!B!.
A similar example can be presented to show that direct release is not allowed either.
For a similarly phrased program with time-transitive flows there is no fixed policy for which the program is interfering, hence time-transitive flows are allowed.


\subsubsection{Bisimulation (BI) based conditions}

The non-disclosure property by Boudol and Matos~\cite{Boudol:Matos:On} as well as the original Flow Locks security condition by Broberg and Sands~\cite{Broberg:Sands:ESOP06} are defined in terms of \emph{bisimulations}.
Bisimulations are used in other security conditions~\cite{Castellani:Boudol:TCS02,Smith:CSFW01,Smith:CSFW03}, however the bisimulation in the presence of dynamic policies are of a somewhat different form called `strong' bisimulation, originally introduced by Sabelfeld and Sands~\cite{Sabelfeld:Sands:CSFW00} to phrase a security condition for concurrent applications.

As we only aim to classify the various security conditions along our facets, we do not present the exact conditions from~\cite{Boudol:Matos:On,Broberg:Sands:ESOP06}.
Instead, we present strong bisimulation in the general setting that we use in this paper that preserves the essence of the cited properties and is classified by our facets in the same way.

%

The purpose of strong bisimulation is to define a relation~$\sim$ between programs.
Two programs $(P_1, P_2) \in \sim$ iff the following holds.
For all $S_1$ with $\langle P_1, S_1 \rangle \overarrow{e_1}{} \langle P_1', S_1' \rangle $, let $f$ be the flow relation when event $e_1$ was produced.
Then for every $S_2$ that is equivalent to $S_1$ as observed by an attacker $A$ under flow relation $f$,
there should exist an event trace $t_2$ such that $\langle P_2, S_2 \rangle \overarrow{t_2}{} \langle P_2', S_2' \rangle $ with $\obs{e_1}$ = $\obs{t_2}$ and $(P_1', P_2') \in \sim$.

Textually, this relates two programs such that if one produces an observation this can be matched by the other, provided that they started with equivalent stores.
In addition, the programs from the resulting configurations should be in the same relation to each other, i.e.\ we require that $P_2'$ can continue to simulate $P_1'$.

A program is then considered secure if it is bisimilar to itself: $P \sim P$.
A program is considered secure if it is bisimilar to itself: $P \sim P$.
An important property of \emph{strong} bisimulation is that it quantifies over equivalent stores \emph{per} observation.
Effectively this means that it considers combinations of memory stores and commands which are impossible in a single-threaded system.
For example, for the command \lstinline!if (x > 0) { y := x}! bisimulation is required for the subcommand \lstinline!y := x! for stores where \lstinline!x! $ \leq 0$.

Since bisimulation is required on all sub-commands regardless of previous flows, security conditions based on strong bisimulation do not allow for (strong) replaying of flows nor direct release.
Time-transitive flows are allowed, because the condition ignores the transitive dependencies and considers for the second part of the transitive flows only those stores that agree on the value that was released in the first part.

\subsubsection{Epistemic security conditions}

\paragraph{Askarov and Sabelfeld (Gradual Release)~\cite{Askarov:Sabelfeld:Gradual}}
The use of epistemic formulations in security conditions for information flow was first found in the Gradual Release property by Askarov and Chong~\cite{Askarov:Sabelfeld:Gradual}.
The policy language contains two security levels \textit{Low} and \textit{High}.
The programming language includes a declassification primitive to allow information flows from \textit{High} to \textit{Low}, explicitly marking the resulting observations as `release events'.

The security condition is then that the knowledge of a \textit{Low} observer should not increase on observations that are not release events:
That is, for all configurations $\langle P, S \rangle$ producing a trace $t\cdot{}e$:
\[
  \neg\mathit{release}(e) \Longrightarrow \ek(P,S,t) = \ek(P,S,t\cdot{}e)
\]
Here, the knowledge of an attacker includes both knowledge of the \textit{Low}-labelled part of the initial store (which as for \Rx{} we denote as $S|_{\it Low}$) as well as what can be learned from the observations produced by the program:
\begin{align*}
  & \ek(P,S,t) = \\
  & \quad \{ S' \mid S|_{\it Low} \not= S'|_{\it Low} \} \mathop{\cup} \\
  & \quad \{ S' \mid \neg \exists t'. \langle P, S' \rangle \overarrow{t'}{} \text{ and } {\it obs}_{\it Low}(t) = {\it obs}_{\it Low}(t') \}
\end{align*}

Gradual Release is only defined for a setting with two security levels and there appears to be no natural extension for multiple levels without making determining choices on allowing transitive flows.
We therefore classify this facet as not available for gradual release.

The Gradual Release definition does not involve an actual dynamic policy as such.
Hence in order to classify the definition along the facets they need to be phrased using the declassification primitive instead of policy change.
For example, replaying flows can be exemplified with:
\begin{minipage}{\columnwidth}
\begin{lstlisting}
a := declassify(b)
a := b
\end{lstlisting}
\end{minipage}
As the first observation is a release event, the requirement that the observer's knowledge stays the same applies only on the second observation.
Having already observed the value of \lstinline!b! in the first observation, the second observation does not teach the observer anything new: replaying flows is allowed.

The concept of Gradual Release can be used to denote more fine-grained policies specifying \emph{what} information may be revealed at each release event, see e.g.~\cite{Askarov:Sabelfeld:Tight,Balliu:2011:ETL}.
Such extensions do not change how the facets are addressed.

As the observations are modelled as assignments, the property does not allow for direct release.

\paragraph{Banerjee et al. (Flowspecs)~\cite{Banerjee+:Expressive}}

Banerjee et al. introduce a modular approach to statically enforcing secure information release.
Standard type-checking for non-interference is employed on most of the program code except for those parts marked as declassifying.
These declassifications need to agree with a \emph{flowspec}, a specification of what information may be released and under which circumstances.
Program verification is used to verify that for each declassification there indeed exists a flowspec that allows for this release.

The enforced security condition, \emph{conditioned gradual release}, is highly similar to gradual release except that for each releasing observation there has to be a flowspec which allows for this release.
This additional expressiveness in the specification of policies has no impact on the classification in our facets, but can provide more guarantees than a gradual release policy.
We discuss the expressivity of policy specification mechanisms in detail in Section~\ref{sec:pollang:expressivity}.

\paragraph{Broberg and Sands (Paralocks)~\cite{Broberg:Sands:PLAS09,Broberg:Sands:Paralocks}}
The Flow Locks policy language and its successor Paralocks have similar security properties to which we refer collectively as Paralocks security.
In the Paralocks policy language, information flows to actors are guarded by \emph{locks} which can be opened and closed.
That is, the flow relation depends on the set of open locks: the more locks are open, the more flows are permitted.

In the Paralocks policy language, security labels take the form $\{ a : L_1, \dots, L_n \}$, denoting that information with this label can be observed by actor $a$ provided that locks $L_1, \dots, L_n$ are open.
The set of open locks is referred to as the lock state.
In contrast, the label $\{a : L_1\}$ is more permissive in any lock state, as it requires less locks to be open.
However, if locks $L_2, \dots, L_n$ are open, both labels are equally permissive.
The Paralocks semantics thus defines a relation $l_1 \sqsubseteq_{LS} l_2$, specifying whether label $l_2$ is at least as restrictive as $l_1$ in lock state $LS$.

Based on gradual release, Paralocks security also defines attacker knowledge as the combination of what can be observed from the initial memory store and what is learned additionally from observations produced by the program.
An attacker $A$ consists of a pair of actor $a$ (the observer) and a set of locks called \emph{capabilities} $C$ (this attacker can observe as if these locks were open).
\begin{align*}
  & \ek(P,S,t) = \\
  & \quad \{ S' \mid S \not\equiv_A S' \} \mathop{\cup} \\
  & \quad \{ S' \mid \neg \exists t'. \langle P, S' \rangle \overarrow{t'}{} \text{ and } \obs{t} = \obs{t'} \}
\end{align*}
Here, $\equiv_A$ considers two stores equivalent if they agree on values with labels $l \sqsubseteq_{\emptyset} \{ a : C\}$, i.e. when no locks are open.

Paralocks security is then defined as follows.
Let $t \cdot e$ be a trace produced by $\langle P, S \rangle$ with $LS$ the lock state at the moment event~$e$ was produced.
For each attacker $A$, consisting of actor $a$ and capability $C$, if $LS \subseteq C$, then it should hold that $\ek(P, S, \obs{t \cdot e}) = \ek(P, S, \obs{t})$.

As identified for the previously discussed epistemic properties, the definition of knowledge means that Paralocks security necessarily allows for replaying flows since this knowledge does not increase when the same information flows again.
On the other hand, the property does not allow direct release as the attacker's observations are only on store changes.

Let $A$ be an attacker observing the second observation resulting from a time-transitive flow.
If the locks necessary for this flow exceed $A$'s capability, the security condition directly considers this secure for $A$.
Otherwise, $A$ would necessarily also have made the first observation of the time-transitive flow, making this effectively a replay for $A$ which we already argued as secure.

\paragraph{Balliu~\cite{Balliu2013}}

In his work, Balliu shows how trace-based conditions such as separability and generalised non-interference fit in a generic security condition and how they can be characterised in epistemic temporal logic.
This condition is of a format similar to the property discussed in Section~\ref{sec:semantics:epistemic}.
\[
  \ek_A(P,\obs{t \cdot e}) \setminus \ek_A(P,\obs{t}) \subseteq \rNow
\]
The security condition is parametric in the release policy $\rNow$, and therefore does not fix all of the facets until this release policy is instantiated.
The definition of knowledge \emph{does} have a fixed definition, (simplified to our setting):
\[
  \ek(P,t) = \{ S' \mid \neg \exists t' . \langle P, S' \rangle \overarrow{t'}{} \text{ and }  \obs{t} = \obs{t'} \}
\]
That is, the attacker excludes all states which cannot produce a trace that consists of the same observations for $A$.
Once again, the fixed definition of knowledge implies that $\ek_A(P,\obs{t \cdot e}) \setminus \ek_A(P,\obs{t}) = \emptyset$ when event $e$ is the replay of earlier flows, and replays are permitted.

We illustrate how this security condition may both permit and deny direct release and time-transitive flows depending on the instantiation of the release policy.

Let $f$ be the current flow relation when event $e$ was produced.
To disallow either facet, the release policy can be specified to only allow the attacker $A$ to exclude those initial stores that are not equivalent on the memory locations observable to $A$.
In particular we do not involve the produced event trace $t$ by the program.
Here, $\equiv^f_A$ considers two stores equivalent if they agree on the values which according to $f$ may be observed by $A$.
\begin{align*}
  \rNow = \{ S' \mid S' \not\equiv^{f}_A S \}
\end{align*}

To allow for time-transitive flows, we allow the attacker to exclude initial states based on observations on levels which the current policy allows to flow to~$A$:
Here, ${\it obs}_A(t,f)$ preserves all events in $t$ that according to $f$ may be observed by $A$.
\begin{align*}
  & \rNow = \\
	& \quad \{ S' \mid \neg \exists t' .  \langle P, S' \rangle \overarrow{t'}{} \text{ and }  {\it obs}_A(t,f) = {\it obs}_A(t',f) \}
\end{align*}

Finally, a release policy which allows for the direct release of information can be achieved by allowing the attacker to exclude stores based on information that was at some point observable to $A$.
Let $\vec{f}$ be the set of all flow relations that were active at any of the events in $t \cdot{} e $.
The release policy can then be defined as:
\[
  \rNow = \{ S' \mid S' \not\equiv^{f}_A S \text{ for some } f \in \vec{f} \}
\]
This allows for direct release though is more permissive than necessary since revocations of flow relations are ignored.

\paragraph{Askarov and Chong~\cite{askarov2012}}

The security condition by Askarov and Chong, by intention, does \emph{not} necessarily allow for replaying information flows.
As for Balliu, the security property is like  in Section~\ref{sec:semantics:epistemic}, except that $\rNow$ is fixed to the set of stores that do not appear equivalent to the attacker $A$ under the flow relation when the last event $e$ was produced:
\begin{align*}
  \rNow = \{ S' \mid S' \not\equiv^{f}_A S \}
\end{align*}
As observed by Buiras and van Delft, this makes the security condition disallow time-transitive flows~\cite{slio}.
Neither does the condition allow for direct release, since observations are based on produced outputs.

\begin{figure}
  \centering
  \begin{tikzpicture}[->,auto,node distance=2cm,semithick]
    \tikzstyle{every state}=[fill=none,draw=black,text=black]

    \node[initial,state] (S)                                {$q_0$};
    \node[state]         (A) [right of=S]           {$q_f$};
    \node[state]         (B) [above right=0cm and 1cm of A] {$q_1$};
    \node[state]         (C) [below right=0cm and 1cm of A] {$q_2$};

    \path (A) edge [above, pos=0.4] node {1} (B)
              edge [below, pos=0.4] node {2} (C)
          (S) edge [bend left]      node {1} (A)
              edge [bend right]     node {2} (A);
  \end{tikzpicture}
  \caption{An automaton modelling a forgetful attacker who only remembers the second value observed (in the domain $\{1,2\}$), in the style of Askarov and Chong~\protect\cite{askarov2012}.}
  \label{fig:forgetful}
\end{figure}
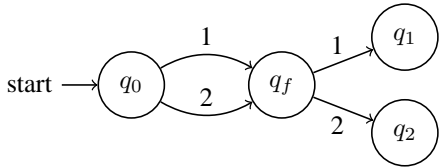

What sets this security condition apart is its definition of attacker knowledge.
Rather than assuming an attacker with perfect recall, Askarov and Chong allow attackers to `forget' (parts of) earlier observations, hence possibly perceiving two different traces as equal.
An attacker $A$ is modelled as a combination of a level and an automaton which makes transitions based on the observed values.
As an example, the automaton in Figure~\ref{fig:forgetful} models an attacker who remembers the second output but forgets the value of the first observation.

Knowledge is now defined as the set of initial stores that an attacker excludes because they cannot produce a trace $t'$ such that the automaton ends in the same state as when observing the actual trace $t$.
Consider an attacker with the automaton from Figure~\ref{fig:forgetful} observing the output on level $A$ of the following program (assuming that ${\tt b} \in \{1,2\}$):
\begin{lstlisting}
  A *\mayFlow* B
b := a
  A *\mayNotFlow* B
b := a
\end{lstlisting}
Only after the second observation does this attacker learn the value of \lstinline!b!, that is at a time when this is \emph{not} allowed by the current flow relation.
Hence, the forgetful attacker model is an effective way to rule out replays of previous information flows.
Replays can still be permitted if only the perfect-recall attacker is considered.

Unfortunately, as identified by Askarov and Chong, demanding security against \emph{all} possible forgetful attackers could be argued unreasonable as it includes `wilfully stupid' and unrealistic attackers.
Van Delft et al.~\cite{delft2015} identify a definitive set of attackers for a progress-insensitive version of the security condition, but identifying a reasonable set for the progress-sensitive version remains an open question.

\subsection{Directions for Future Research}
\label{sec:semantics:questions}


It could be useful to formulate a generic security property which is modular in its facet
classification.
Such a property would allow the designer to make a choice for each facet, and then simply achieve this by enabling or disabling the necessary components in the generic property
(although not all combination of facets may be possible, as the existing properties in Table~\ref{table:classification} suggests).
The security property by Balliu exhibits some of this genericity, but lacks identifiable components for each facet.




We consider it possible that our set of facets is incomplete and additional facets of security properties will be identified in the future.
For each new facet, we recommend a similar treatment to survey existing security conditions and identify what influences its classification.


When the security condition determines the facets, all information releases are treated equally.
An interesting generalisation  would be to let the program  explicitly state the \emph{intended facets per release}.
For example, a program could annotate some releases as ``permanent'', indicating to the security property that these may be replayed (as opposed to the unannotated releases).
The security property has to treat such specified intentions properly.\footnote{One could argue that \Rx{} already presents such a system.
By placing code in a transaction the programmer indicates the intention to disallow time-transitive flows within this code block.
However, as argued in the discussion of \Rx{}, these intentions are ignored by the security condition which still treats all releases as allowing for time-transitive flows.}

\section{Classifications of Dynamic Policies}
\label{sec:pollang}

In this section we take a closer look at the anatomy of policy schemes and policy specification mechanisms,
with the intention of introducing a standard terminology for discussing and comparing them.
We discuss the different \emph{control levels} at which a scheme can operate, and how
these levels taken together, in combination with assignment of labels in a system,
form what we generally refer to as a policy. 
We further use this anatomy as an
overlay to shed new light on the ``dimensions of declassification'' introduced by
Sabelfeld and Sands \cite{Sabelfeld:Sands:JCS}, and resolve some perceived unclarities and
ambiguities among the dimensions. We categorise existing policy specification mechanisms in 
the literature accordingly.
Finally we derive a framework for reasoning about the kinds of \emph{invariants} a
mechanism can express in a generated policy scheme, as a starting point for comparing 
expressiveness between different mechanisms. 


\subsection{The anatomy of dynamic policies}
\label{sec:pollang:anatomy}

\newcommand{\orderings}{\ensuremath{\mathcal{F}}}
\newcommand{\dynamicpol}{\ensuremath{\delta}}
\newcommand{\metapol}{\ensuremath{\mu}}

A dynamic policy scheme can be understood, discussed and classified in terms of a \emph{hierarchy of control}, 
consisting of the following control levels:
\begin{itemize}
  \item Level 0 control -- \orderings{}, a \emph{set} of possible \emph{flow relations} between the information
        sources and sinks available in the system.
  \item Level 1 control -- \dynamicpol{}, a \emph{determining function} selecting which flow relation in \orderings{} 
        should be active. We refer to the arguments to this function as the \emph{discriminator}.
  \item Level 2 control -- \metapol{}, a \emph{meta policy} controlling the way in which the current flow relation 
        may be changed.
\end{itemize}
The control levels allow us to be explicit about what it is that makes a policy dynamic: 
the possibility to define a determining function \emph{and} have an input to this function
that changes during program execution.
Arguably, one could have a meta-meta policy which in turn controls the meta policy.
However, with no loss of generality we group all further abstractions in the 
\emph{meta policy} class, since the meta policy can also be taken to control itself.

As a concrete example, consider a scheme consisting of two security labels, \emph{Secret} and \emph{Public}.
Potentially the set \orderings{} could contain all four possible flow relations that involve these 
two security labels, but in this example it contains only two:
one in which \emph{Public} information may flow to \emph{Secret} but not vice-versa; 
and one in which information may also flow from \emph{Secret} to \emph{Public}.
The former flow relation is the default, and the latter is available when using
a special ``declassify'' operator.
The determining function \dynamicpol{} then decides whether information may be released from 
\emph{Secret} to \emph{Public} or not, based on the current value of its argument, $S$. In
this example, $S$ can range over boolean values, indicating whether ``declassify'' is used
or not. The meta-policy simply always allows transitions back and
forth between true and false.


Formally we characterise the determining function as:
\[
  \dynamicpol{} : S \rightarrow \orderings{}
\]
Here $S$ is the information used to determine which flow relation is currently active.
$S$~could range over boolean values as in our simple example; it could be partial information from the current program state, as is the case in Paragon~\cite{Paragon} or the work by Chong and Myers~\cite{Chong:Myers:CCS04}.
$S$~may also consist of other information such as lexical location in the source code (as is done by Boudol and Matos~\cite{Boudol:Matos:On}) or `asynchronous' information external to the program (e.g. the acts-for hierarchy among principals used by Hicks et al.~\cite{Hicks+:Dynamic}).

In turn, the meta policy controls the changes between flow relations caused by the determining function.
Strictly speaking, a meta policy aims to control the transitions between flow relations, and does not
require the determining function as a ``proxy''. For example, a meta policy could, again based on some 
(meta) information $M$ such as program state, impose a constraint on the progress of flow relations:
\[
  \metapol{} : M \rightarrow 2^{\orderings{} \times \orderings{}}
\]
Here the pair $(f_1, f_2) \in \orderings{} \times \orderings{}$ indicates that transition from flow 
relation $f_1$ to $f_2$ is permitted. In this way, it would be easy to for example define a \metapol{}
specifying the meta policy that the flow relation between information only becomes more liberal.
However in the surveyed literature, we find that the meta policy is typically defined in terms of
controlling how the information $S$ used by \dynamicpol{} may change, rather than the resulting
flow relation determined by \dynamicpol{} -- simply an extra level of indirection.
That is, a meta policy instead typically has the characterisation:
\[
  \metapol{} : M \rightarrow 2^{S \times S}
\]
Note that the determining function does not solely serve as a level of indirection for the meta policy:
the meta policy specifies how the current flow relation \emph{may} be changed, but it is the determining function that specifies what the current flow relation \emph{is}.

To demonstrate that the control hierarchy functions as a terminology for policy specification mechanisms, 
we show how various mechanisms from literature fit onto these levels:
\begin{itemize}

  \item The programming language Jif~\cite{Myers:POPL99} has a declassification function which can be
        described as temporarily changing $S$, making the intent clear such that \dynamicpol{} provides
        a flow relation which allows for the declassification.
        The relation between declassification and dynamic policies is discussed in more detail in 
        Section~\ref{sec:declassification}.
        The decision to declassify is restricted by components such as authority and declassification
        robustness~\cite{Zdancewic:Myers:CSFW01}, which constitute $M$. Both \dynamicpol{} and \metapol{}
        are pre-determined.
        The flow relations in \orderings{} are determined by the labels used, in combination with the
        \emph{acts-for} hierarchy. If changes to the acts-for hierarchy are allowed, 
        as per Hicks et al \cite{Hicks+:Dynamic}, then this hierarchy is also included in $S$.

  \item The programming language Paragon~\cite{Paragon} allows for the specification of Paralocks policies (see Section~\ref{sec:semantics:survey}) in a Java-like language.
        Hence the current flow relation is determined by the lock state, 
        which constitutes~$S$. In turn, these locks are information in the program state and are protected by
        locks themselves\footnote{We observe that the possibility to place locks on policies is part of the 
        specification mechanism used in Paragon, but not included in the Paralocks specification language.}, 
        making that second set of locks determine the meta-policy, i.e. when the first set of locks can be 
        opened and closed. The labels include specification of how they are interpreted with respect to
        changes in the lock state, meaning that the set of flow relations, and the behaviour of \dynamicpol{}
        (as well as \metapol{}) can be customised for each specific policy scheme.

  \item The programming language \Rx{}~\cite{Swamy+:Managing} incorporates the \textit{RT} 
        framework~\cite{Li:Mitchell:Winsborough:Design} to specify a flow relation among \emph{roles}.
        The active flow relation is determined by the set of memberships and delegations specified on
        each role, which form~$S$.
        Roles also carry labels which specify who can observe the current members of a role, hence
        forming the set $M$ determining on which secret data the decision to change flow relation
        (add or remove memberships and delegations) may depend.
  
  \item Matos and Cederquist~\cite{Matos2014} present a security condition for distributed computations 
        (Distributed Non-interference). The default flow relation between security labels can be relaxed
        using the lexical construct from earlier work on the non-disclosure property~\cite{Boudol:Matos:On},
        making $S$ the locality in the code.
        These scopes with more liberal flow relations may only be entered when the node on which the 
        computation runs allows for it. Each node specifies its own regulation on the allowed added flows,
        making $M$ the locality in the network.
  
\end{itemize}

We note that the apparently clear separation offered by the levels of control is not necessarily mirrored 
in actual specification mechanisms. As noted, in Paragon each data source and sink is annotated with a
security label that specifies not only a static behaviour, but also how that label should be interpreted
in different lock states. In other words, the information on what the flow relations are and how they are
determined is distributed across all the labels in a program, not cleanly as a single determining function.
This does not mean that Paragon could not still be understood and described in terms of the levels of control.

Another observation is that the higher levels of control, \dynamicpol{} and \metapol{}, have two components
where control can be exercised: in the definition of \dynamicpol{}, resp. \metapol{}, and in the argument
to the respective function. To contrast these two possibilities, \Rx{} allows control over the argument
provided to \metapol{} (who can observe the members of a role), but \metapol{} itself is fixed.
This is opposed to the meta-policy by Matos and Cederquist where the argument to \metapol{} is fixed to
be the node on which the computation runs, but \metapol{} itself can be defined by the policy designer.

\subsection{Rethinking the Dimensions of Declassification}
\label{sec:pollang:reclassify}

Looking at the literature, it is clear that the four ``dimensions of declassification'' introduced
by Sabelfeld and Sands \cite{Sabelfeld:Sands:JCS} -- ``what'', ``who'' ``where'' and ``when'' --
are significantly different from each other in nature.
In particular aspects of the ``what'' dimension are largely orthogonal from the other three,
while many uses of ``where'' and ``when'' often coincide. There are also
different aspects grouped within the same dimension that are so disparate as to be incomparable.
For example, the ``where'' dimension intends to cover both \emph{code locality} and \emph{level locality},
which are only remotely related at best.

We propose that these dimensions should be discussed for each of our identified dynamic policy levels 
\emph{individually}. We identify that not every dimension is relevant for every policy level, as 
summarised in Table~\ref{tbl:decldimensions}.
This insight resolves some of the confusion in the declassification dimensions,
showing that by taking the anatomy of a policy into account while discussing its 
``declassification' dimensions, we arrive at a clearer framework for discussing and comparing 
security conditions.

We present a short summary of each of the declassification dimensions and discuss them with respect to the policy anatomy.

\begin{table}
\begin{center}
  \begin{tabular}{@{}|l|*{4}{c}|@{}}
    \cline{2-5}
    \multicolumn{1}{ c|  }{} & What & Who & Where & When \\ 
    \hline
    \orderings{} & + & - &  Level locality only & - \\
    \dynamicpol{} & - & + & + & + \\
    \metapol{} & - & + & + & + \\ \hline
  \end{tabular}
\end{center}
\vspace*{0.7ex}
\caption{Revisiting the declassification dimensions; + indicates that a dimension concern can be addressed by that policy component, - that it cannot.}
\label{tbl:decldimensions}
\end{table}

\para{What} -- 
Policies can dictate that only parts of a secret may be released (e.g. the last digits of a credit card). 
In addition this dimension covers quantitative release which is better characterised by ``how much'' and can be achieved using an information-theoretic approach (e.g. \cite{Clark+:ENTCS}).
Although the decision to e.g. increase the amount of information that may be released comes from different components, the possibility to \emph{express} this dimension only exists naturally in the ordering between information itself, i.e.
when specifying flow relations.

\para{Who} -- 
This dimension is concerned with being able to express who controls the release.
In particular, it is sensible to prevent an attacker from abusing the release mechanism, 
as is the motivation for robust declassification~\cite{Zdancewic:Myers:CSFW01}.
Since the decision to declassify can be controlled both by the determining function 
and the meta policy, this dimension can be addressed on either level.
By nature this dimension talks about control over flow relations, and therefore 
is not relevant on the level of the flow relation itself.

\para{Where} --
This dimension is split into two different forms of locality: \emph{level} locality and \emph{code} locality.

Level locality addresses the concern where information may flow relative to the security levels of the system.
This dimension is particularly present in \emph{intransitive non-interference}~\cite{Rushby:92}, which is exemplified by a policy which allows information to flow from security level {\it Secret} to {\it Declassify} and from {\it Declassify} to {\it Public}, but not directly from {\it Secret} to {\it Public}.
This can be expressed in a flow relation using downgrading relations (Mantel~\cite{Mantel:FME01}),
but could also be addressed by the \dynamicpol{} and \metapol{} controls on the ordering.
The latter is achieved by changing between flow relations such that only one of the two flows is
permitted at any specific time (this essentially requires \emph{time-transitive flows},
see Section~\ref{sec:semantics:dimensions}).

Code locality allows policies to describe where syntactically in the code information may be released.
One could sort of view this as level locality except the information should not pass through the 
{\it Declassify} level but through a lexical declassification construct in the program's code.
Similar for the Who dimension, code locality is concerned with controlling which flow relation is active, 
and can therefore only be addressed by \dynamicpol{} or \metapol{}.

\para{When} --
A policy can dictate that information may only be released after (or before) a certain time has passed.
This temporal restriction can be based on various elements, such as real time, the size of the secret or relative to other events in the system.
Although the original presentation of this dimension splits it into various classes, all temporal controls need to be addressed by \dynamicpol{} and \metapol{} as they concern the decision to change the ordering of information.

%
%
%
%

When we now reclassify policy mechanisms by the levels first and the dimensions second, 
the classification becomes clear and unambiguous.
We briefly show this for the examples used in Section~\ref{sec:pollang:anatomy}.
\footnote{None of the considered examples addresses the ``what'' dimension, or 
support intransitive flows in the flow relations, thus we do not discuss this dimension further.}

For Jif, the discriminator for the determining function \dynamicpol{} is given by a combination
of the declassification construct, which concerns the ``where'' dimension (both code and level locality), and
the acts-for hierarchy, which concerns the ``who'' dimension.
As a meta-policy, authority provides a meta control on the decision to
declassify in the ``who'' dimension.
Robustness does so as well, and in addition partially addresses the ``what'' dimension by limiting what information can be declassified.
For Paragon, both \dynamicpol{} and \metapol{} are regulated by locks which concern the ``when'' 
dimension: information flows are allowed relative to the actions of opening and closing locks.
Paragon also has a lexically scoped version of opening a lock, which works in the ``where'' dimension
(code locality).
Implementations in Jif and Paragon can combine the programming and policy language to encode
requirements in other dimensions~\cite{Askarov:Sabelfeld:ESORICS2005,Paragon,paragontut},
but these are not a natural part of the policy language.

Both \dynamicpol{} and \metapol{} in \Rx{} use the ``who'' dimension: who is a member of a role
determines the active flow relation, and who can view the current members of a role decides what
policy change can be made.
The security framework for distributed non-interference by Matos and Cederquist finds a fit in
the ``where'' dimension for \dynamicpol{} as the lexical flow construct determines where in the code the additional
flows are allowed. The meta policy also fits in that dimension, as it determines where in the network each flow construct is allowed.

\subsection{The expressiveness of policy languages}
\label{sec:pollang:expressivity}

Different policy specification mechanisms offer a variety of expressiveness, from the
simplest fixed two-level systems, up to full policy specification \emph{languages}
like those found in Jif \cite{jif}, \Rx{}~\cite{Swamy+:Managing} or Paragon \cite{Paragon}.
We can have intuitive ideas regarding the relative expressiveness of such mechanisms,
but what are the measures by which we can compare them formally? In this section we speculate
on how a formal framework for comparison could be constructed.


Montagu et al \cite{Montagu13} construct a framework for comparing ``label models'',
or \emph{policy schemes} in our terminology. We argue that such an approach is too simple
for comparing expressiveness of full policy specification languages.

Consider a policy scheme which has three labels called $\text{TopSecret}$, $\text{Secret}$ and $\text{Public}$.
      By default, the flow relation consists of the three flows 
      $\text{Public} \rightarrow \text{Secret}$, $\text{Public} \rightarrow \text{TopSecret}$
      and $\text{Secret} \rightarrow \text{TopSecret}$, so the labels form a strict hierarchy
      of security levels. The policy scheme also allows data to be declassified from $\text{Secret}$
      to $\text{Public}$. When declassifying, the flow relation is then the same three flows from before,
      with $\text{Secret} \rightarrow \text{Public}$ added. These are the only two flow relations
      possible. We refer to this scheme as TSP.

A second scheme has three labels simply called $A$, $B$ and $C$. Any of the six possible
    flows between two labels can be allowed or not independently of other flows. In other words,
    all $2^6$ conceivable flow relations involving these three labels are possible, and a
    programmer can freely change between them. We refer to this scheme as ABC.

A first 
attempt at comparing expressiveness could look at the possibility to \emph{embed} one
\emph{policy scheme} in the other\footnote{This is the comparison done by Montagu et al \cite{Montagu13}.}.
That is, the embedding scheme should contain at least the same set of flow relations as the embedded scheme.
In this case we could embed TSP in ABC using
$\text{TopSecret} = A$, $\text{Secret} = B$ and $\text{Public} = A$, and use the corresponding
two matching flow relations. We could then claim that the second is at least as
expressive as the first, by virtue of having at least as many labels allowing at least
the same flow relations. However, such an attempt misses an important aspect of expressiveness.
If we were to use ABC in place of TSP, what (other than regimen) stops us from making
one of the ``other'' flow relations active? In particular, we have no guarantees that we will
not use a flow relation in which $A$, proxying as $\text{TopSecret}$, can flow to other labels.
ABC is certainly more \emph{flexible} than TSP -- but when embedding, added
flexibility is not a good thing. Restrictions matter!

For a policy scheme, the degree of flexibility is already fixed, so there is never any room
for expressiveness. Truly, expressiveness should be compared at the level of policy specification
mechanisms. Consider mechanisms $PSM_1$ and $PSM_2$: we have that $PSM_1$ is at least as expressive as 
$PSM_2$ if, for every possible policy scheme that $PSM_2$ can generate, $PSM_1$ can generate a scheme that
can embed it -- \emph{including restrictions}. But how can we express restrictions formally?

The examples used so far show the need for restrictions at the level of what flow relations are
possible. However, not all such restrictions are equally important. For our example above, the
fact that when using ABC we could end up in contexts where the flow relation allows \emph{fewer}
flows than any of the ones matching those of TSP, is arguably acceptable -- the system would
still be secure. But the fact that we could end up in a context where $A$, representing $\text{TopSecret}$,
can flow at all is not acceptable, as it means that using ABC we cannot give the same security
guarantees \emph{by construction}. Hence, what matters is the ability to express
\emph{invariants} over flow relations -- specifically, invariants that concern the \emph{absence}
of some (set of) flows.

In this work we identify two principal forms of invariants that we want the ability to express.
The first form are the \emph{invariants over sets of flow relations}. Such invariants can be
global, for example ``no flow from $\text{TopSecret}$ to any other level is ever allowed'';
or \emph{conditional}, for example ``flows from $\text{Secret}$ to $\text{Public}$ are not
allowed, except when declassifying''. The second principal form are the
\emph{invariants over sequences of flow relations}. A simple example could be the
Gradual Release property \cite{Askarov:Sabelfeld:Gradual} that the policy may only
change to become more liberal over time. Another more complicated example is a strong
Chinese Wall property stating that if a flow $\text{CompanyOne} \rightarrow X$ has ever
been allowed at any point, then $X \rightarrow \text{CompanyTwo}$ may not be allowed at
any point in the future~\cite{brewer1989chinese}. 

To formalise these notions we first observe that given some starting state $S_0$ and the set of possible
transitions as given by the range of $\mu$\footnote{If we also know how the argument $M$ to $\mu$ may change
over time, we can have better precision than considering the whole range of $\mu$. As argued, this could
be accomplished using yet another level of meta-policy that governs how $M$ may change, or baking this into
$\mu$ itself.}, we can enumerate all possible sequences
of discriminators by iteratively applying all possible transitions. Let $\Sseq$
be the set of all such sequences. A global invariant over sets of reachable flow relations is then a property $\Phi$
such that 
\[
  \forall{S_0 \cdot \ldots \cdot S_n} \in \Sseq . \Phi(\delta(S_0)) \land \ldots \land \Phi(\delta(S_n))
\] holds. A \emph{conditional} invariant adds a filter $\Psi$ such that
\begin{multline*}
  \forall{S_0 \cdot \ldots \cdot S_n} \in \Sseq . \\
    [\Psi(S_0) \Rightarrow \Phi(\delta(S_0))] \land \ldots \land [\Psi(S_n) \Rightarrow \Phi(\delta(S_n))]
\end{multline*}
holds.  An invariant over \emph{sequences} of flow relations is a property $\Phi$ such that
\[
  \forall{S_0 \cdot \ldots \cdot S_n} \in \Sseq . \Phi(\delta(S_0) \cdot \ldots \cdot \delta(S_n))
\] holds. We can easily imagine a conditional version of invariants over sequences too, with a similar filter
based on the domain of $\mu$, however we have not identified any compelling examples.

\begin{framed}
\centerline{\bf Comparing expressiveness of policy languages}
\noindent
A policy specification mechanism $PSM_1$ is at least as restrictive
as another mechanism $PSM_2$, if for every possible policy scheme that $PSM_2$ can generate,
$PSM_1$ can generate a scheme that can embed it, including enforcing the same (negative)
invariants.
\end{framed}

The kinds of invariants we have categorised here capture the majority of all conceivable invariants
that we may want a policy scheme to enforce, however, there are more complex invariants that
cannot be expressed in these terms. Our invariants are essentially \emph{safety properties}, 
and not all desired invariants can be expressed in terms of these. 
Our framework of invariants should thus be seen as a starting point for
formalising comparisons between policy specification mechanisms, not a completed journey,
and the formalisation of further points in the space of invariants is an open research question.

\section{Relation to Declassification}
\label{sec:declassification}

The term \emph{declassification} (or more generally \emph{downgrading})
has long been used to signify the deliberate
change of security label on data, to allow it to be used more liberally than
before. This is not specific to research on information flow, or even
security in general -- the term is used with this meaning outside of 
technical contexts as well.

For information flow specifically, however, the exact meaning of the term
is not clear. It has been used, we argue, to refer also to things that are
not aligned with the natural definition given above. 

We will first argue for proper uses of the term declassification, and
consequently also for what we consider mis-uses of the term. We will then
go on to discuss, within the proper uses, different meanings that can
be given to the term; specifically we identify two different \emph{flavours}
of declassification, discuss their distinctive differences, and relate
these to the facets from section \ref{sec:semantics:dimensions}. 


\subsection{The term ``declassification''}

For terminology, we want to establish a clear difference between the use
of the term ``declassification'' (or ``downgrading'') and dynamic policy
change. In technical terms, mechanisms that achieve declassification can
be treated as specific uses of dynamic policies, 
but the concepts are not equivalent. The important distinction we want to make 
is that
declassification is, by nature, \emph{data-centric}, as opposed to affecting
a policy. It is possible to declassify a piece of data, but it is \emph{not}
possible to ``declassify'' a policy. A policy can be changed to become more
(or less) liberal -- data can be declassified and then \emph{used} more
liberally.
Further, it seldom makes sense to talk about ``declassifying'' a program or
a computation\footnote{
Unless one is literally revealing a previously secret piece of source code.}
-- instead that program or computation can be said to run \emph{under a more
liberal policy}.

Historically this distinction has not been entirely clear, leading to some confusion about specifically which aspects of a policy specification mechanism or semantic
property that are referred to when talking about the declassification taking
place in a system. In particular 
the survey on ``Dimensions and principles of declassification'' by Sabelfeld
and Sands\cite{Sabelfeld:Sands:JCS} uses the term ``declassification''
as an umbrella 
to include everything related
to dynamic policies -- and even some mechanisms which are purely
static orderings. 
We now hope to foster a more fine-grained use of the terminology.

\subsection{Flavours of declassification}

In the literature we identify two distinctly different flavours of
the (proper) use of the term: \emph{relabelling} and \emph{copying release}.
\para{Relabelling} 
In classic military terms, declassification corresponds to the physical operation of 
changing the security classification of a document. This process 
is mirrored in some systems that declassify data
by effectively replacing its label with a more liberal one.
We refer to this as the \emph{relabelling} approach to declassification. 
In information flow
systems, this behaviour can be 
simulated by 
changing a sufficiently fine-grained
global policy in such a way that it puts less restriction on the usage of
the data in question. Gradual Release \cite{Askarov:Sabelfeld:Gradual} is an example of such a policy. 
It should be clear that this use of the term can be seen as a direct
instance of dynamic policies, where the policy is made successively more
liberal.

\para{Copying Release}
The other flavour of the term declassification is the systems that use a
specific declassifying \emph{operator}\footnote{More generally \emph{operation} --
it does not need to be an operator per se, even if that is the most common case.}
that allows a single exceptional flow that would otherwise violate the
prevailing policy. In effect, declassifying in this sense creates a \emph{copy}
of the original data\footnote{More generally, the result of an expression whose result depends on the original data.}
available under a more liberal label.
We refer to this as the \emph{copying release} approach.
Systems with this form of declassification include Jif \cite{jif} and
JOANA \cite{joana14it}.
Semantic properties for such systems are typically phrased in terms of a flow
against the normal ordering being allowed specifically if the operation
causing it is a distinguished declassification operation.

We can also view such operators as instances of dynamic policy change, where an
application of the operator corresponds to a sequence of operations in which:
the global policy is temporarily weakened, the flow happens, and the policy
is restored to its previous state. This is how declassification is typically encoded in e.g. Paragon \cite{Paragon}. Note that this works in a sequential setting; when
we introduce concurrency one would need to prevent concurrent threads from exploiting the temporary policy change, for example by making the sequence of operations atomic.

\begin{table}
  \centering
  \begin{tabular}{c}
  \begin{tabular}{@{}|l|*{4}{c}|@{}}
    \cline{2-5}
    \multicolumn{1}{ c|  }{} & T  & R & D & W \\ \hline
    Relabelling
      & + & + & +/-  & +\\
    Copying release
      & + & - & +/-  & +\\
	  \hline
	\end{tabular} \vspace{5pt}\\
	  T: time-transitive, R: replay, D: direct release, W: whitelisting
  \end{tabular}
\vspace*{1ex}	
  \caption{Necessary facets for a semantic property accommodating different declassification flavours.}
	\label{table:facets:declassification}
\end{table}

Interestingly, a semantic property intended to accommodate such operations is
restricted in the choices that can be made for the various facets introduced
in Section \ref{sec:semantics:dimensions}.
Table~\ref{table:facets:declassification} summarises the compatibility of the facits with the two flavours of declassification.

The information made available by a copying release is intended to be persistent, but not permanent.
The operator effectively releases a \emph{copy} of the information to a location with a different security level. 
This makes the release persistent, since the information can now be accessed from that security level instead, and the original classification of the copy is forgotten.
As a consequence an accompanying security property needs to allow for both time-transitive flows and whitelisting. The release is, however, not permanent, as the released information is intended to be accessed \emph{only} from the location containing the copy.
If the declassified copy of the data is deleted, it is simply no longer available under the liberal label.
This demands that the security property does not allow (weak) replaying of flows.

The only facet (of the ones we discuss in this paper) that is not fixed when employing a declassification operator, is in the treatment of direct release.
Direct release can be allowed \emph{only} if the policy is fine-grained enough to distinguish between each possible data item to be declassified.
In practice, if the result of arbitrary expressions can be declassified, direct release would not be feasible.


\section{Concluding discussion}
We reiterate that our aim has been to synthesise knowledge about dynamic policies,
with the purpose to increase understanding and help facilitate future work within
the domain of information flow control.
Our anatomy of policy schemes can give would-be authors of policy specification
mechanisms a better understanding of the nature of what they propose, and the
tools to sharpen it to achieve the desired expressiveness.
Would-be creators of programming languages and systems incorporating information
flow policies can draw inspiration and understanding from our discussion on the
nature of declassification, and further look to our facets to make conscious
choices regarding the nature of their security properties, to ensure that
they truly capture the desired degree of security.
And interested researchers can draw inspiration from the less illuminated
and understood corners and areas that we leave uncovered or identify as
open research questions.
In short, we believe that the foundations laid down in this paper will make
future work on information flow control sharper, easier, and stronger.

\para{Acknowledgments}
This paper benefited from the comments of
Musard Balliu,
Pablo Buiras,
Owen Arden,
Sebastian Hunt,
Andrei Sabelfeld,
and the anonymous reviewers.
This work is partly funded by the Swedish funding agencies SSF and VR.


\bstctlcite{IEEEexample:BSTcontrol}
\bibliographystyle{IEEEtranS}
\bibliography{literature}

\begin{thebibliography}{10}
\providecommand{\url}[1]{#1}
\csname url@samestyle\endcsname
\providecommand{\newblock}{\relax}
\providecommand{\bibinfo}[2]{#2}
\providecommand{\BIBentrySTDinterwordspacing}{\spaceskip=0pt\relax}
\providecommand{\BIBentryALTinterwordstretchfactor}{4}
\providecommand{\BIBentryALTinterwordspacing}{\spaceskip=\fontdimen2\font plus
\BIBentryALTinterwordstretchfactor\fontdimen3\font minus
  \fontdimen4\font\relax}
\providecommand{\BIBforeignlanguage}[2]{{%
\expandafter\ifx\csname l@#1\endcsname\relax
\typeout{** WARNING: IEEEtranS.bst: No hyphenation pattern has been}%
\typeout{** loaded for the language `#1'. Using the pattern for}%
\typeout{** the default language instead.}%
\else
\language=\csname l@#1\endcsname
\fi
#2}}
\providecommand{\BIBdecl}{\relax}
\BIBdecl

\bibitem{Boudol:Matos:On}
A.~{Almeida Matos} and G.~Boudol, ``On declassification and the non-disclosure
  policy,'' in \emph{Proc. IEEE Computer Security Foundations Workshop}, 2005,
  pp. 226--240.

\bibitem{Matos2014}
A.~{Almeida Matos} and J.~Cederquist, ``{Distributed Noninterference},'' in
  \emph{Euromicro Int. Conf. on Parallel, Distributed, and Network-Based
  Processing}.\hskip 1em plus 0.5em minus 0.4em\relax IEEE Computer Society,
  2014, pp. 760--764.

\bibitem{Askarov:Sabelfeld:ESORICS2005}
A.~Askarov and A.~Sabelfeld, ``Security-typed languages for implementation of
  cryptographic protocols: {A} case study,'' in \emph{Proc. European Symp. on
  Research in Computer Security}, ser. LNCS, vol. 3679.\hskip 1em plus 0.5em
  minus 0.4em\relax Springer-Verlag, 2005.

\bibitem{Askarov:Sabelfeld:Gradual}
A.~Askarov and A.~Sabelfeld, ``Gradual release: Unifying declassification,
  encryption and key release policies,'' in \emph{Proc. IEEE Symp. on Security
  and Privacy}, May 2007, pp. 207--221.

\bibitem{askarov2012}
A.~Askarov and S.~Chong, ``{Learning is change in knowledge: Knowledge-based
  security for dynamic policies},'' in \emph{Computer Security Foundations
  Symposium (CSF), 2012}.\hskip 1em plus 0.5em minus 0.4em\relax IEEE, 2012,
  pp. 308--322.

\bibitem{Askarov:Sabelfeld:Tight}
A.~Askarov and A.~Sabelfeld, ``{Tight Enforcement of Information-Release
  Policies for Dynamic Languages},'' in \emph{IEEE Computer Security
  Foundations Symposium}, 2009, pp. 43--59.

\bibitem{Balliu2013}
M.~Balliu, ``A logic for information flow analysis of distributed programs,''
  in \emph{Secure IT Systems}.\hskip 1em plus 0.5em minus 0.4em\relax Springer
  Berlin Heidelberg, 2013, vol. 8208, pp. 84--99.

\bibitem{Balliu:2011:ETL}
M.~Balliu, M.~Dam, and G.~Le~Guernic, ``Epistemic temporal logic for
  information flow security,'' in \emph{Proceedings of the ACM SIGPLAN 6th
  Workshop on Programming Languages and Analysis for Security}, ser. PLAS
  '11.\hskip 1em plus 0.5em minus 0.4em\relax ACM, 2011, pp. 6:1--6:12.

\bibitem{Banerjee+:Expressive}
A.~Banerjee, D.~Naumann, and S.~Rosenberg, ``Expressive declassification
  policies and modular static enforcement,'' in \emph{Proc. IEEE Symp. on
  Security and Privacy}.\hskip 1em plus 0.5em minus 0.4em\relax IEEE Computer
  Society, 2008, pp. 339--353.

\bibitem{Castellani:Boudol:TCS02}
G.~Boudol and I.~Castellani, ``Non-interference for concurrent programs and
  thread systems,'' \emph{Theoretical Computer Science}, vol. 281, no.~1, pp.
  109--130, Jun. 2002.

\bibitem{brewer1989chinese}
D.~F. Brewer and M.~J. Nash, ``The chinese wall security policy,'' in
  \emph{Security and Privacy, 1989. Proceedings., 1989 IEEE Symposium
  on}.\hskip 1em plus 0.5em minus 0.4em\relax IEEE, 1989, pp. 206--214.

\bibitem{Broberg:Sands:ESOP06}
N.~Broberg and D.~Sands, ``Flow locks: Towards a core calculus for dynamic flow
  policies,'' in \emph{Programming Languages and Systems. 15th European
  Symposium on Programming, ESOP 2006}, ser. LNCS, vol. 3924.\hskip 1em plus
  0.5em minus 0.4em\relax Springer Verlag, 2006.

\bibitem{Broberg:Sands:PLAS09}
N.~Broberg and D.~Sands, ``Flow-sensitive semantics for dynamic information
  flow policies,'' in \emph{ACM SIGPLAN Fourth Workshop on Programming
  Languages and Analysis for Security (PLAS 2009)}.\hskip 1em plus 0.5em minus
  0.4em\relax Dublin: ACM, June 15 2009.

\bibitem{Broberg:Sands:Paralocks}
N.~Broberg and D.~Sands, ``Paralocks -- role-based information flow control and
  beyond,'' in \emph{Symposium on Principles of Programming Languages
  (POPL)}.\hskip 1em plus 0.5em minus 0.4em\relax ACM, 2010.

\bibitem{Paragon}
N.~Broberg, B.~van Delft, and D.~Sands, ``{Paragon for Practical Programming
  with Information-Flow Control},'' in \emph{Programming Languages and
  Systems}, ser. LNCS, 2013, vol. 8301, pp. 217--232.

\bibitem{slio}
P.~Buiras and B.~van Delft. {Dynamic Enforcement of Dynamic Policies -
  Technical Report}. \url{http://slio.bitbucket.org/slio-tr.pdf}. Accessed:
  2015-02-11.

\bibitem{Chong:Myers:CCS04}
S.~Chong and A.~C. Myers, ``Security policies for downgrading,'' in \emph{{ACM}
  Conference on Computer and Communications Security}, Oct. 2004, pp. 198--209.

\bibitem{Clark+:ENTCS}
D.~Clark, S.~Hunt, and P.~Malacaria, ``Quantitative analysis of the leakage of
  confidential data,'' in \emph{QAPL'01, Proc. Quantitative Aspects of
  Programming Languages}, ser. ENTCS, vol.~59.\hskip 1em plus 0.5em minus
  0.4em\relax Elsevier, 2002.

\bibitem{Clark:Hunt:FAST08}
D.~Clark and S.~Hunt, ``{Non-Interference for Deterministic Interactive
  Programs},'' in \emph{Proc. 5th International Workshop on Formal Aspects in
  Security and Trust (FAST2008)}, ser. Lecture Notes in Computer Science.\hskip
  1em plus 0.5em minus 0.4em\relax Springer-Verlag, 2008.

\bibitem{Cohen:InformationA}
E.~S. Cohen, ``Information transmission in computational systems,'' \emph{ACM
  SIGOPS Operating Systems Review}, vol.~11, no.~5, pp. 133--139, 1977.

\bibitem{Focardi:Gorrieri:Classification}
R.~Focardi and R.~Gorrieri, ``A classification of security properties for
  process algebras,'' \emph{J. Computer Security}, vol.~3, no.~1, pp. 5--33,
  1995.

\bibitem{Goguen:Meseguer:Noninterference}
J.~A. Goguen and J.~Meseguer, ``Security policies and security models,'' in
  \emph{Proc. IEEE Symp. on Security and Privacy}, Apr. 1982, pp. 11--20.

\bibitem{Halpern:2008}
\BIBentryALTinterwordspacing
J.~Y. Halpern and K.~R. O'Neill, ``Secrecy in multiagent systems,'' \emph{ACM
  Trans. Inf. Syst. Secur.}, vol.~12, no.~1, pp. 5:1--5:47, Oct. 2008.
  [Online]. Available: \url{http://doi.acm.org/10.1145/1410234.1410239}
\BIBentrySTDinterwordspacing

\bibitem{Hicks+:ACSAC06}
B.~Hicks, K.~Ahmadizadeh, and P.~D. McDaniel, ``From languages to systems:
  Understanding practical application development in security-typed
  languages,'' in \emph{ACSAC}.\hskip 1em plus 0.5em minus 0.4em\relax IEEE
  Computer Society, 2006.

\bibitem{Hicks+:Dynamic}
M.~Hicks, S.~Tse, B.~Hicks, and S.~Zdancewic, ``Dynamic updating of
  information-flow policies,'' in \emph{Foundations of Computer Security
  Workshop}, Jun. 2005, pp. 7--18.

\bibitem{Li:Mitchell:Winsborough:Design}
N.~Li, J.~Mitchell, and W.~Winsborough, ``{Design of a role-based
  trust-management framework},'' in \emph{IEEE Symposium on Security and
  Privacy}, 2002, pp. 114--130.

\bibitem{Mantel:FME01}
H.~Mantel, ``Information flow control and applications---{Bridging} a gap,'' in
  \emph{Proc. Formal Methods Europe}, ser. LNCS, vol. 2021.\hskip 1em plus
  0.5em minus 0.4em\relax Springer-Verlag, Mar. 2001, pp. 153--172.

\bibitem{Montagu13}
B.~Montagu, B.~C. Pierce, and R.~Pollack, ``A theory of information-flow
  labels,'' in \emph{Proceedings of the 2013 IEEE Computer Security Foundations
  Symposium}, Jun. 2013.

\bibitem{Myers:POPL99}
A.~C. Myers, ``{JF}low: Practical mostly-static information flow control,'' in
  \emph{Proc. ACM Symp. on Principles of Programming Languages}, Jan. 1999, pp.
  228--241.

\bibitem{ML00}
A.~C. Myers and B.~Liskov, ``Protecting privacy using the decentralized label
  model,'' \emph{ACM Transactions on Software Engineering and Methodology},
  vol.~9, no.~4, pp. 410--442, 2000.

\bibitem{jif}
A.~C. Myers, L.~Zheng, S.~Zdancewic, S.~Chong, and N.~Nystrom, ``{Jif}: {J}ava
  information flow,'' Jul. 2001--2006, software release.
  \url{http://www.cs.cornell.edu/jif}.

\bibitem{Preibusch:POLICY2011}
S.~Preibusch, ``Information flow control for static enforcement of user-defined
  privacy policies,'' in \emph{Policies for Distributed Systems and Networks
  (POLICY), 2011 IEEE International Symposium on}.\hskip 1em plus 0.5em minus
  0.4em\relax IEEE, 2011, pp. 133--136.

\bibitem{Rushby:92}
J.~M. Rushby, ``Noninterference, transitivity, and channel-control security
  policies,'' SRI International, Tech. Rep. CSL-92-02, 1992.

\bibitem{Sabelfeld:Sands:CSFW00}
A.~Sabelfeld and D.~Sands, ``Probabilistic noninterference for multi-threaded
  programs,'' in \emph{Proc. IEEE Computer Security Foundations Workshop}, Jul.
  2000, pp. 200--214.

\bibitem{Sabelfeld:Sands:CSFW05}
A.~Sabelfeld and D.~Sands, ``Dimensions and principles of declassification,''
  in \emph{Proc. IEEE Computer Security Foundations Workshop}, Jun. 2005, pp.
  255--269.

\bibitem{Sabelfeld:Sands:JCS}
A.~Sabelfeld and D.~Sands, ``Declassification: Dimensions and principles,''
  \emph{Journal of Computer Security}, vol.~15, no.~5, pp. 517--548, 2009.

\bibitem{Smith:CSFW01}
G.~Smith, ``A new type system for secure information flow,'' in \emph{Proc.
  IEEE Computer Security Foundations Workshop}, Jun. 2001, pp. 115--125.

\bibitem{Smith:CSFW03}
G.~Smith, ``Probabilistic noninterference through weak probabilistic
  bisimulation,'' in \emph{Proc. IEEE Computer Security Foundations Workshop},
  2003, pp. 3--13.

\bibitem{joana14it}
G.~Snelting, D.~Giffhorn, J.~Graf, C.~Hammer, M.~Hecker, M.~Mohr, and
  D.~Wasserrab, ``Checking probabilistic noninterference using joana,''
  \emph{it - Information Technology}, vol.~56, pp. 280--287, Nov. 2014.

\bibitem{Stoughton+:PLAS14}
A.~Stoughton, A.~Johnson, S.~Beller, K.~Chadha, D.~Chen, K.~Foner, and
  M.~Zhivich, ``You sank my battleship!: A case study in secure programming,''
  in \emph{Proceedings of the Ninth Workshop on Programming Languages and
  Analysis for Security}, ser. PLAS'14.\hskip 1em plus 0.5em minus 0.4em\relax
  ACM, 2014, pp. 2:2--2:14.

\bibitem{Swamy+:Managing}
N.~Swamy, M.~Hicks, S.~Tse, and S.~Zdancewic, ``{Managing Policy Updates in
  Security-Typed Languages},'' in \emph{Proceedings of the 19th IEEE Workshop
  on Computer Security Foundations}, 2006.

\bibitem{Swamy:2008}
N.~Swamy and M.~Hicks, ``{Verified Enforcement of Stateful Information Release
  Policies},'' in \emph{Proceedings of the Third ACM SIGPLAN Workshop on
  Programming Languages and Analysis for Security}, ser. PLAS '08.\hskip 1em
  plus 0.5em minus 0.4em\relax ACM, 2008, pp. 21--32.

\bibitem{paragontut}
B.~van Delft, N.~Broberg, and D.~Sands, ``{Programming with Paragon},'' in
  \emph{Proc. 2013 Marktoberdorf Summer School}, ser. NATO Science Series,
  2013.

\bibitem{delft2015}
B.~van Delft, S.~Hunt, and D.~Sands, ``{Very Static Enforcement of Dynamic
  Policies},'' in \emph{Principles of Security and Trust}.\hskip 1em plus 0.5em
  minus 0.4em\relax Springer, 2015.

\bibitem{Zakinthinos:SP97}
A.~Zakinthinos and E.~Lee, ``A general theory of security properties,'' in
  \emph{In Proceedings of the IEEE Symposium on Security and Privacy}.\hskip
  1em plus 0.5em minus 0.4em\relax Society Press, 1997, pp. 94--102.

\bibitem{Zdancewic:Myers:CSFW01}
S.~Zdancewic and A.~C. Myers, ``Robust declassification,'' in \emph{Proc. IEEE
  Computer Security Foundations Workshop}, Jun. 2001, pp. 15--23.

\bibitem{Zhang2012}
C.~Zhang, ``{Conditional Information Flow Policies and Unwinding Relations},''
  in \emph{Trustworthy Global Computing}, ser. Lecture Notes in Computer
  Science.\hskip 1em plus 0.5em minus 0.4em\relax Springer, 2012, vol. 7173,
  pp. 227--241.

\bibitem{zheng2005dynamic}
L.~Zheng and A.~C. Myers, ``Dynamic security labels and noninterference,'' in
  \emph{Formal Aspects in Security and Trust}.\hskip 1em plus 0.5em minus
  0.4em\relax Springer, 2005, pp. 27--40.

\end{thebibliography}

\appendices
\section{Glossary of terminology}

\newcommand{\glosterm}[2]{\item[{\bf #1}] \hfill \\ #2 \\[-0.5em]}

\begin{description}

  \glosterm{Copying release}{Making a copy of (derived) data available under a more liberal \emph{security label}.}

  \glosterm{Declassification}{The deliberate change of the \emph{security level} on data to allow it to be used more liberally.}

  \glosterm{Determining function}{Function that, based on its arguments (the \emph{discriminator}), determines a \emph{flow relation}.}

  \glosterm{Dimension (of declassification)}{A classifying axis on the basis of the \emph{declassification} goal (what, where, when, or to whom information may flow).}

  \glosterm{Direct release}{A \emph{Facet}; information is considered released as soon as the current flow relation allows it to flow.}

  \glosterm{Discriminator}{Argument to the \emph{determining function}.}

  \glosterm{Downgrading}{See \emph{Declassification}.}

  \glosterm{Dynamic policy}{\emph{Information flow policy} under which the \emph{flow relations} may change during computation.}

  \glosterm{Facet}{An aspect of a \emph{security condition} that determines whether a particular class of information flows is accepted as secure.}

  \glosterm{Flow relation}{Relates components in the program or system between which information is permitted to flow, e.g. as an ordering between \emph{security labels}.}	

  \glosterm{Exclusion Knowledge}{The set of secrets that could not have produced a given observation.}

  \glosterm{Hierarchy of control}{Division of a \emph{policy scheme} in three levels of control, each level controlling the one below it.}

  \glosterm{Information flow policy}{Specification of the information flows permitted during program execution.}

  \glosterm{Knowledge}{The set of all secrets that could have produced a given observation.}

  \glosterm{Meta policy}{Specification of the way in which the current \emph{flow relation} may be changed.}

  \glosterm{Noninterference}{\emph{Security condition} that defines absence of information flow, typically by saying that changing a secret input will not cause a change in public outputs.}

  \glosterm{Policy}{See \emph{information flow policy}.}

  \glosterm{Policy scheme}{A set of \emph{flow relations} and transition between them, in isolation from any particular program or system.}

  \glosterm{Policy specification mechanism}{Mechanism or language to construct a \emph{policy scheme}.}

  \glosterm{Relabelling}{Replacing the \emph{security label} on information, placing less restrictions on the usage of that information.}
  
  \glosterm{Release Policy}{Determines what information may be released with each observation, possibly based on various aspects such as the current flow relation when the observation was produced, the attacker, or the produced observation.}

  \glosterm{Replaying flows}{A \emph{Facet}; information that has flowed previously can flow again, regardless of what the flow relation dictates.}

  \glosterm{Security condition}{Semantic specification of when a program satisfies a given security policy.}

  \glosterm{Security label}{A label attached to particular parts of the program or system in order to express security concerns.}

  \glosterm{Static policy}{\emph{Information flow policy} under which the \emph{flow relation} may not change during computation.}

  \glosterm{Termination insensitivity}{A \emph{Facet}; when information that is revealed by the termination or output progress of an application is always permitted by the \emph{security condition}.}

  \glosterm{Time-transitive flows}{A \emph{Facet}; information from \emph{security level} $A$ may flow to level $C$ via a third level $B$, while there is no moment in time where the \emph{flow relation} allows flows from $A$ to $C$ directly.}

  \glosterm{Whitelisting flows}{A \emph{Facet}; an information flow is allowed whenever some part of the \emph{flow relation} permits it.}

\end{description}

\end{document}